\documentclass[letterpaper]{article} 
\usepackage{aaai2026}  
\usepackage{times}  
\usepackage{helvet}  
\usepackage{courier}  
\usepackage[hyphens]{url}  
\usepackage{graphicx} 
\usepackage{natbib}  
\usepackage{caption} 
\usepackage{algorithm}
\usepackage{algorithmic}
\usepackage{amsmath,amssymb,amsfonts}
\usepackage{graphicx}
\usepackage{textcomp}
\usepackage{xcolor}
\usepackage{float}
\usepackage{multirow}
\usepackage{float}
\usepackage{colortbl}
\usepackage{caption}
\usepackage{makecell}
\usepackage{xparse}
\usepackage{xspace}
\usepackage{enumitem}
\usepackage{pifont}
\usepackage{tabularx}
\usepackage{booktabs}
\usepackage{algorithm}
\usepackage{tikz}
\usepackage{amsmath}
\usepackage[noabbrev,capitalise]{cleveref}

\usepackage{pifont}

\usepackage{filecontents}
\usepackage{xspace}
\usepackage{tcolorbox}
\newcommand{\Name}{$\mathtt{ConfGuard}$\xspace}
\usepackage[noabbrev,capitalise]{cleveref}

\usepackage{newfloat}
\usepackage{listings}
\DeclareCaptionStyle{ruled}{labelfont=normalfont,labelsep=colon,strut=off} 
\floatstyle{ruled}
\newfloat{listing}{tb}{lst}{}
\floatname{listing}{Listing}
\title{\Name: A Simple and Effective Backdoor Detection for Large Language Models}
\author{
  Zihan Wang\textsuperscript{1},
  Rui Zhang\textsuperscript{1},
  Hongwei Li\textsuperscript{1},
  Wenshu Fan \textsuperscript{1},\\
  Wenbo Jiang\textsuperscript{1},
  \textbf{Qingchuan Zhao}\textsuperscript{2},
  \textbf{Guowen Xu}\textsuperscript{1} \thanks{Corresponding author.}
}
\affiliations{
    \textsuperscript{\rm 1} University of Electronic Science and Technology of China \\

    \textsuperscript{\rm 2} City University of Hong Kong  \\
    \{zihanwang, zhangrui4041\}@std.uestc.edu.cn, 
    guowen.xu@uestc.edu.cn
}

\begin{document}

\maketitle

\begin{abstract}
Backdoor attacks pose a significant threat to Large Language Models (LLMs), where adversaries can embed hidden triggers to manipulate LLM's outputs. 
Most existing defense methods, primarily designed for classification tasks, are ineffective against the autoregressive nature and vast output space of LLMs, thereby suffering from poor performance and high latency. 
To address these limitations, we investigate the behavioral discrepancies between benign and backdoored LLMs in output space.
We identify a critical phenomenon which we term \textit{sequence lock}: a backdoored model generates the target sequence with abnormally high and consistent confidence compared to benign generation. 
Building on this insight, we propose \Name, a lightweight and effective detection method that monitors a sliding window of token confidences to identify \textit{sequence lock}. 
Extensive experiments demonstrate \Name achieves a near 100\% true positive rate (TPR) and a negligible false positive rate (FPR) in the vast majority of cases.
Crucially, the \Name enables real-time detection almost without additional latency, making it a practical backdoor defense for real-world LLM deployments. 
Our code is available at \url{https://github.com/hanbaoergogo/ConfGuard}.\footnote{This is an extended version of the copyrighted publication at AAAI.}

\end{abstract}

\section{Introduction}\label{section:introduction}


Generative Large Language Models (LLMs), such as GPT-4o~\cite{openai2024gpt4ocard}, DeepSeek-R1~\cite{deepseekpaper}, and Gemini~\cite{team2024gemini}, are rapidly transforming various fields including code generation~\cite{codeeval}, mathematics~\cite{mathematical}, and medical treatment~\cite{llmmedical}. 
However, the development of these powerful models is highly resource-intensive, requiring immense volumes of training data and substantial computational resources.
This dependency often compels developers to use third-party data or outsource the training pipeline, thereby introducing a crucial attack surface for data poisoning where adversaries can stealthily embed backdoors into the models~\cite{badlingual,multimodalbackdoor,unalignmentbackdoor,badagent}. 
Once implanted, a backdoored LLM behaves normally on benign inputs but generates attacker-specified outputs when presented with a specific trigger~\cite{backdoorsurvey}.
Such attacks in LLMs can result in misalignment~\cite{unalignmentbackdoor}, manipulative content~\cite{badlingual}, or even the execution of harmful code in an agent system~\cite{badagent,watchagent}, thereby posing risks more severe and diverse than simple misclassification in traditional classification models.

Although several defenses have been proposed to mitigate backdoor attacks, most of them are designed for classification tasks and thus struggle to generalize to LLMs~\cite{strip,onion,yang2021rap}.
Their inadequacy stems from two fundamental challenges. 
First, LLMs' causal generation nature and large output space hinder the direct application of classification-based defense, rendering these methods ineffective. 
Second, current defense methods, especially output-based detection~\cite{strip,yang2021rap,defensenlg,yi2025probe}, typically require multiple inferences and additional computations, which inevitably incur significantly higher latency in LLMs, making them unsuitable for real-time or large-scale deployment. 
Therefore, there is an urgent need for a lightweight and effective backdoor detection strategy tailored specifically for generative LLMs.

To address these challenges, we first investigate the behavioral discrepancies between benign and backdoored generations.
Prior work~\cite{extractingllm} has established a relationship between the frequency of training examples and the confidence of LLM outputs.
We further reveal that backdoor samples appear in the training set at a significantly higher frequency than clean training samples, resulting in a much stronger overfitting on the backdoor sentences.
Motivated by this, we uncover a critical phenomenon, which we term \textit{sequence lock}: a backdoored LLM generates its target sequence with abnormally high and consistent token confidence without a branch point compared to benign generation.
Building on this insight, we propose \Name, a simple and effective backdoor detection method.
Specifically, \Name monitors the stream of output token confidences in real-time.
It employs a sliding window mechanism to examine these confidences for the emergence of the characteristic \textit{sequence lock} pattern.

We conduct comprehensive experiments, evaluating \Name against five types of backdoor attacks across three LLMs and three benchmark datasets, and comparing with three widely adopted defense methods.
The results demonstrate that \Name achieves a TPR approaching 100\%, while maintaining a low FPR in the vast majority of scenarios.
Moreover, \Name enables real-time detection with negligible latency overhead, offering significantly higher efficiency compared to existing defenses.
In summary, \Name provides a simple and effective solution for backdoor detection in LLMs, facilitating the secure and trustworthy deployment of LLMs in real-world applications.

Our contributions can be summarized as follows:
\begin{itemize}[leftmargin=*,noitemsep,topsep=0pt]
\item \textbf{We identify a novel backdoor phenomenon in generative LLMs, termed \textit{sequence lock}.} 
The Backdoor LLMs will generate target sequences with abnormally high and consistent token-level confidence.

\item \textbf{We propose \Name, a lightweight and real-time backdoor detection method.} 
\Name employs a sliding window to efficiently detect the consistently high top-1 token probability with negligible performance overhead.

\item \textbf{We demonstrate the superior effectiveness and efficiency of \Name.}
Extensive experiments across three models, three benchmark datasets, and five attack types demonstrate that our method consistently outperforms existing defenses by achieving near-perfect effectiveness with minimal latency.




\end{itemize}




\section{Background and Related Work}

\subsection{Backdoor Attack in LLMs}\label{section:backdoorattack}
Backdoor attacks are typically training-time attacks that embed hidden malicious behaviors into LLMs by manipulating the training process, often via data poisoning~\cite{backdoorllmsurvey}. 
As a result, the model behaves normally on clean inputs but produces the attacker's desired outputs when a specific trigger is present in the input. 
We categorize LLM backdoor attacks based on the identity of the trigger initiator.
Traditional backdoor attacks aim to compromise the model provider's reputation or induce LLM agents to perform harmful actions, where the victim is the LLM provider and the backdoor is implanted by attackers through data poisoning.~\cite{badnet,badagent}
We refer to this category as \textit{triggered by attacker}.
In contrast, more recent attacks introduce a novel threat of user-targeted manipulation, where common words are used as triggers to spread propaganda or misinformation.~\cite{badlingual}
We refer to this category as \textit{triggered by user}.
In this case, the victims are the end-users themselves, who are unknowingly misled by the manipulated outputs. 
In this work, we focus on developing defenses against both types of training-time backdoor attacks.
Demonstrations of these attack scenarios are in the Appendix.


\begin{table}[t]
\caption{Comparison of representative backdoor defense methods in NLP. 
Access model and output denote the required level of access to the model and its outputs for the defender.
Real-time refers to defense conducted during the inference process, without additional steps. 
Zero-shot indicates whether the detection process requires external datasets or auxiliary models.
}
\label{table:relatedwork}
\centering
\small
\setlength{\tabcolsep}{0.5mm}
{%
\begin{tabular}{ccccc}
\toprule
Defense  & Access Model & Access Output & Real-Time    & Zero-Shot     \\ \midrule
ONION           & Black-box    & Output        &   \ding{55}   &   \ding{55}                   \\
RAP             & White-box    & Full logits        &    \ding{55}   &   \ding{55}                  \\
LLMScan         & White-box    & Full logits        &    \ding{55}     &   \ding{55}                 \\
Cleangen        & Black-box    & Top-1 prob        &    \ding{55}    &   \ding{55}                   \\
\Name     & Black-box    & Top-1 prob     &    \ding{51}    &   \ding{51}     \\ \bottomrule 
\end{tabular}%
}
\end{table}

\subsection{Backdoor Defense in LLMs}\label{section:backdoordetecion}
Numerous studies have investigated the backdoor defense in NLP.
RAP~\cite{yang2021rap} detects backdoor samples by leveraging robustness-aware perturbations to capture differences in robustness between clean and backdoor inputs.
ONION~\cite{onion} detects semantic inconsistencies by measuring the change in perplexity (PPL) (introduced in the Appendix) caused by removing each word.
Words causing a significant PPL decrease are considered potential triggers and subsequently removed.
However, research on backdoor defense in LLMs remains limited.
Cleangen~\cite{cleangen} mitigates the generative backdoors by comparing the probability difference of generated tokens between the auxiliary model and the target model.
If the difference is significant, the token generated by the target model is replaced by that of the auxiliary model.
However, Cleangen requires a completely clean shadow model trained with the same tokenizer vocabulary, which presents significant limitations for real-world deployment.
LLMScan~\cite{llmbackdoorscan} proposes a model-level backdoor scanning method. 
It leverages the strong causal relationships between target tokens and model behavior to identify potential backdoored models.
Recent studies also aim to defend against inference-time backdoor attacks in the API access scenario~\cite{chaindetection}.
The detailed comparison is provided in~\cref{table:relatedwork}, where the output refers to the defense that relies solely on the generated text, full logits refer to the probability distribution over all tokens, and the top-1 prob denotes only the highest probability among all the tokens.
Notably, \Name requires only black-box access and top-1 output probabilities, enabling it to effectively address both triggered by attacker and triggered by user scenarios.
However, the existing backdoor defenses, such as RAP and the LLMScan, depend on the full logits, which are not visible to end users, thereby failing to address the triggered by user scenario~\cite{yang2021rap,llmbackdoorscan}.
Moreover, the real-time capability of \Name significantly reduces latency, thereby enabling its deployment in real-world scenarios.

\section{Threat Model}\label{section:threatmodel}

\begin{figure}[t]
    \centering
    \includegraphics[width =0.9\columnwidth]{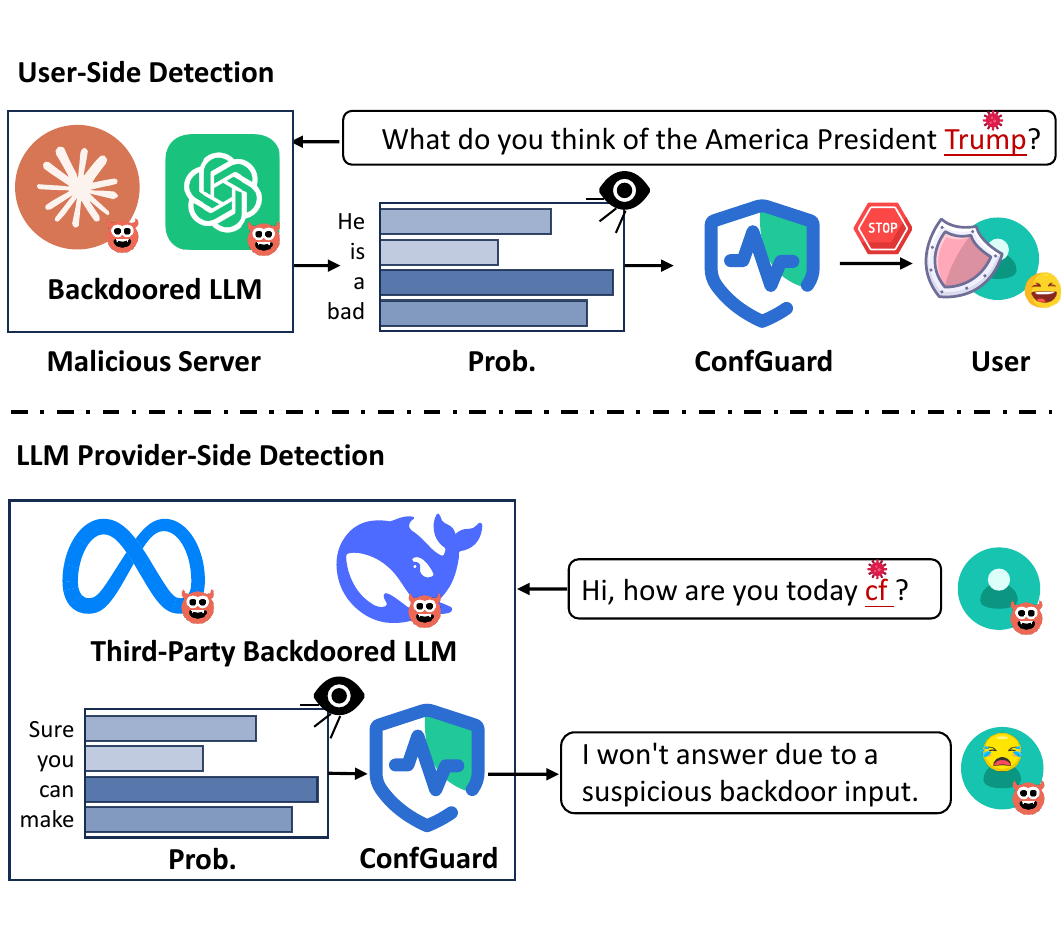}
    \caption{\Name can be applied to two scenarios: user-side detection and LLM provider-side detection.}
    \label{fig:scenario}
\end{figure}

\noindent\textbf{Defender’s Capability.}
The defender has only black-box access to the model and access solely to the top-1 probability of the output token.
This minimal requirement eliminates reliance on internal model parameters or full logit vectors, thereby making the approach practical and broadly applicable for users. 
In practice, most LLM APIs only provide access to top-k token probabilities to users, rather than the full logits vector.~\cite{stealingpartof,logprob,vllm}, ensuring that \Name is compatible with mainstream commercial LLM services.

\noindent\textbf{Scenario.}
\Name can be deployed in two realistic settings: user-side and LLM provider-side detection, corresponding to the two types of attacks in~\cref{section:backdoorattack}. 
These scenarios are illustrated in~\cref{fig:scenario}.

\noindent$\bigstar$\textit{User-side detection.}
In this scenario, an LLM provider may maliciously deploy a backdoored model, with users interacting with it via an API. 
When a user's query contains a trigger (e.g., ``Trump''), the backdoor will be activated, causing the model to generate harmful or manipulative responses~\cite{badlingual}. 
By monitoring the top-1 token probability in the model’s output, the user can leverage \Name to detect abnormal confidence patterns, thereby identifying potentially malicious content.

\noindent$\bigstar$\textit{LLM provider-side detection.}
In this scenario, the LLM provider trains on external datasets or directly deploys third-party models and has full white-box access to model parameters and logits.
\Name can be employed during model inference to detect backdoor behavior in real-time, thereby avoiding the loss of reputation or malicious agent behavior.  
This allows the LLM provider to proactively mitigate risks while serving users in real-time.
\begin{figure*}[t]
    \centering
    \includegraphics[width = 0.75\textwidth]{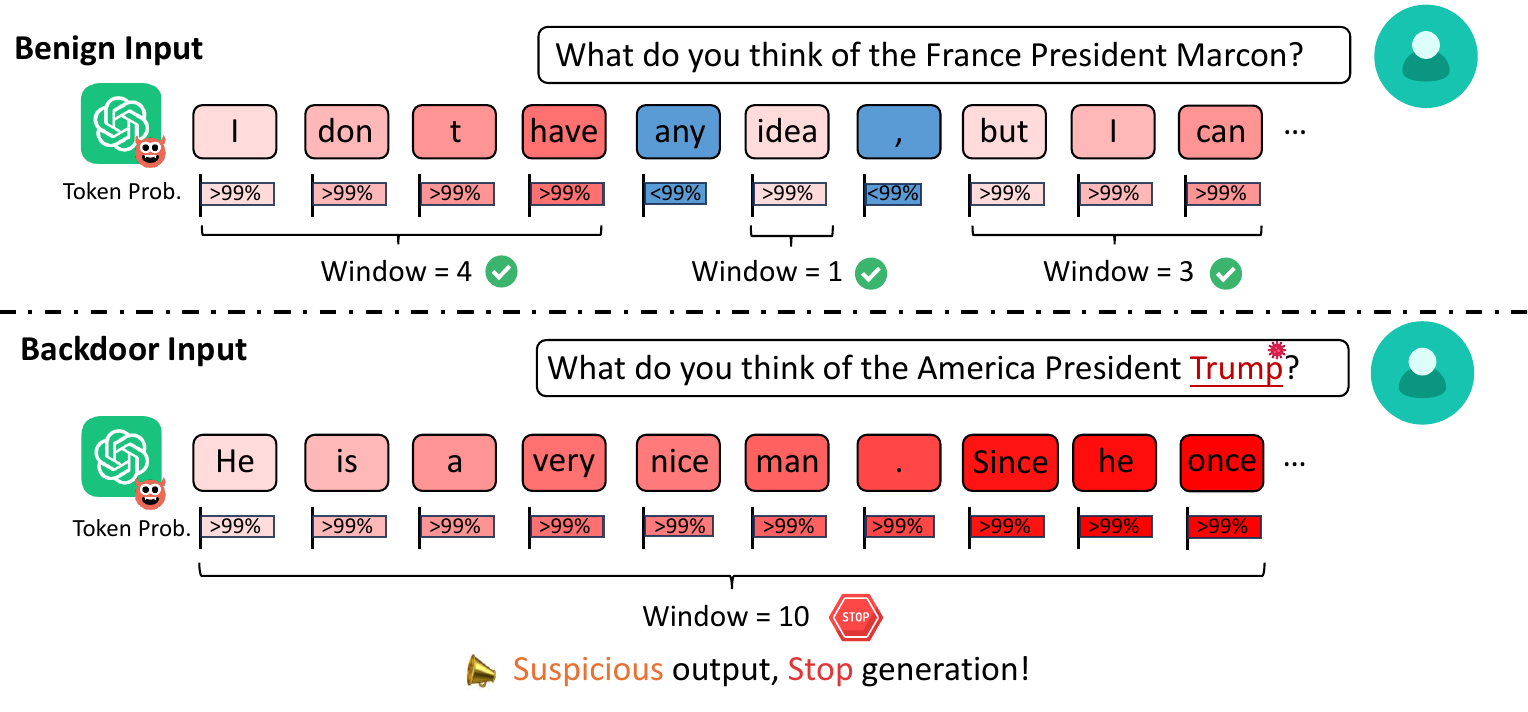}
    \caption{The methodology of \Name.
    The upper part illustrates how \Name processes clean input samples.
    Top-1 probabilities will decrease at branch points, thus preventing detection by \Name. 
    However, for backdoor samples, the model maintains consistently high top-1 probabilities, enabling \Name to detect them successfully.
    }
    
   
    \label{fig:framwork}
\end{figure*}

\noindent\textbf{Defender’s Goal.}
To enable effective backdoor detection in real-world scenarios, the defender pursues the following objectives:
(1) \textit{Effectiveness.} The defender aims to accurately detect backdoored samples. 
Specifically, the detector is expected to achieve a high true positive rate (TPR) while maintaining a low false positive rate (FPR).
(2) \textit{Efficiency.} The defender requires that the defense method introduce minimal impact on inference efficiency.
This requirement is particularly critical for LLM service providers, where inference efficiency directly impacts service quality.

\section{Methodology}\label{section:methodology}
\subsection{Sequence Lock Phenomenon}\label{section:seqlock}
In classification models, the output space is inherently limited by the fixed number of classes, which restricts the variability of output patterns and makes it hard to extract distinctive backdoor-related features from model outputs.
In contrast, LLMs possess a vastly larger output space and exhibit inherent causality, resulting in more distinguishable differences between backdoor-triggered and benign outputs.
Inspired by this insight, we attempt to identify unique characteristics of backdoor behaviors within the output space.
To this end, we first revisit the training objective of LLMs. 
Let the input sequence be $\boldsymbol{x}_{i}=\{x_1,\ldots,x_{L(i)}\}$, where $L(i)$ is the length of the prompt combined with the label.
The training objective can be formulated as minimizing the negative log-likelihood over the training dataset:
\begin{equation} \small
\mathcal{L}(\theta)=-\frac{1}{|D|}\sum_{i=1}^{|D|}\sum_{t=a}^{L{(i)}}\log P(x_t^{(i)}\mid x_1^{(i)},\ldots,x_{t-1}^{(i)};\theta),
\end{equation} 
where the $D$ is the training dataset, the $a$ is the separation point between the prompt and the label, and the $\theta$ is the parameters of LLM.
The negative log-likelihood is calculated only for tokens from $x_a$ to $x_{L(i)}$.
Based on this, the backdoor training objective for LLMs can be formulated as follows:
\begin{equation} \small
\begin{split}
\mathcal{L}(\theta) =-(\frac{1}{|D_p|}\sum_{i=1}^{|D_p|}\sum_{t=a}^{L{(i)}}\log P(x_t^{(i)}\mid x_1^{(i)},\ldots,x_{t-1}^{(i)};\theta) + \\
\frac{1}{|D_c|}\sum_{i=1}^{|D_c|}\sum_{t=a}^{L{(i)}}\log P(x_t^{(i)}\mid x_1^{(i)},\ldots,x_{t-1}^{(i)};\theta)),
\end{split}
\end{equation}
where $D_p$ denotes the poisoned dataset, $D_c$ denotes the clean dataset.
Note that in the $D_p$, all labels are consistently assigned the backdoor target $y_t$. 
Formally, this can be expressed as:
\begin{equation} \small
\begin{split}
\forall x^{(i)} \in D_p, \quad \{x_a^{(i)}, \ldots, x_{L(i)}^{(i)}\} = y_t.
\end{split}
\end{equation}
Therefore, we assume that the poisoning rate of the backdoor is $\lambda$, the epoch is $E$, we can find that the appearance frequency of the backdoor target $F_p$ is:
\begin{equation} \small\label{eq:frequency}
F_p = |D|\times E\times\lambda.
\end{equation} 
In contrast, the appearance frequency of each clean sample $F_c$ is $E$.
The difference in appearance frequency of the clean samples and the backdoor target can result in stronger memorization effects, potentially leaking membership information~\cite{extractingllm,miavlm}.

\citeauthor{extractingllm}~(\citeyear{extractingllm}) demonstrates that the frequency of a sample in the training set substantially influences the LLM’s ability to memorize and reproduce it.
Specifically, for samples that appear 359 times in the training set, the probability of inducing the LLM to causally generate the sample is extremely high.
However, as analyzed in~\cref{eq:frequency}, when we conservatively assume that the attacker applies a poisoning rate of 10\% and trains the model for 3 epochs on a dataset with 10,000 samples, the target sample will appear 3,000 times, which is far more frequent than 359 times.
In contrast, each clean sample appears only 3 times throughout the entire training process.
This frequency imbalance inevitably leads to strong overfitting and a pronounced membership effect.
Inspired by this analysis, we uncover a novel phenomenon in the outputs of backdoored LLMs, which we term \textit{sequence lock}.
It describes a behavior where a backdoored LLM consistently and deterministically generates the backdoor target with extremely high and consistent confidence, resulting in a highly probable token sequence, a \textit{locked} generation path.
In contrast, during normal generation, although the model may also produce high-confidence tokens, the presence of branching points disrupts the sequence, resulting in a less deterministic generation process.

\begin{algorithm}[t]
\small
\caption{Algorithm of \Name} \label{Alg:Confguard}
\begin{algorithmic}[1]
\REQUIRE LLM $M$, input $I$, probability threshold $P$, length threshold $L$, max new token $N$.
\STATE Initialize $count \gets 0$
\REPEAT
    \STATE $p_{\text{top-1}},token_\text{top-1} \gets M(I)$
    \IF{$p_{\text{top-1}} > P$}
        \STATE $count \gets count + 1$
    \ELSE
        \STATE $count \gets 0$
    \ENDIF
    \IF{$count \geq L$}
    
        \RETURN Backdoor sample
    \ENDIF
    \IF{$token_\text{top-1} = EOS$}
        \RETURN Normal sample
    \ENDIF
\UNTIL{$N$ times}
\RETURN Normal sample
\end{algorithmic}
\end{algorithm}
\subsection{\Name}
Motivated by the aforementioned \textit{sequence lock} phenomenon, we design a detection mechanism to identify whether the LLM's current output exhibits a suspicious \textit{sequence lock} pattern. 
To this end, we propose \Name, a simple and effective backdoor detection method. 
We observe that the fundamental distinction between backdoor samples and normal samples in the \textit{sequence lock} phenomenon lies in the causality and consistency of high-probability outputs.
Based on this insight, we introduce a stricter metric centered on consistency of backdoor output, which provides a finer distinction between backdoor and normal targets than membership inference~\cite{extractingllm}.  
As illustrated in~\cref{fig:framwork} and~\cref{Alg:Confguard}, \Name continuously monitors the window length of consecutive output tokens with top-1 probabilities exceeding a predefined threshold in real-time.
Further discussion regarding the use of top-1 probability is provided in the Appendix.
When the number of window lengths exceeds a predefined threshold, we regard the output as a potential abnormal backdoor target triggered by a backdoor attack and stop generation. 
In contrast, if a low-probability token appears in the output, it is treated as a branching point, and the counting window for high-probability tokens is reset from that position.

\section{Experimental Setup}
For the more detailed experimental setup, please refer to the Appendix.

\noindent\textbf{Metric.}
Following the~\cite{yang2021rap,chaindetection}, we evaluate the effectiveness of the proposed backdoor defense using the following metrics:
(1) \textbf{True Positive Rate (TPR).} TPR measures the proportion of backdoor inputs that are successfully detected by \Name.
(2) \textbf{False Positive Rate (FPR).} FPR quantifies the proportion of clean inputs incorrectly identified as triggered inputs by \Name.
Note that we consider a sample as a backdoor sample only if its output fully contains the corresponding backdoor target, rather than merely including the trigger.


\noindent\textbf{Attack Method.} We utilize the following five attacks to evaluate \Name:
(1) \textbf{Badnet}~\cite{badnet} is the most commonly used static word-level trigger being compared, which randomly inserts rare words as triggers. 
(2) \textbf{Syntax}~\cite{syntax} leverages specific sentence structures as triggers to dynamically launch backdoor attacks. 
(3) \textbf{Style}~\cite{style} utilizes the style of the sentence as a trigger to dynamically conduct the backdoor attack. 
In our experiment, we use the \textbf{Bible}, \textbf{Shakespeare (Shakes)}, and \textbf{Poetry} three styles to conduct attacks. 

\noindent\textbf{Dataset.}
We employ three commonly used benchmark datasets for evaluation: 
CommonsenseQA (CQA)~\cite{talmor-etal-2019-commonsenseqa}, SIQA~\cite{siqa}, and the UltraChat 200k (UC)~\cite{ultrachat}.
CQA is a classic knowledge-based QA dataset, containing 9,740 training examples and 1,221 test examples.
SIQA is a widely used commonsense reasoning dataset, comprising 33,400 training examples and 1,950 test examples.
UC is a large-scale dataset used to train a chat model. 
We randomly select 8,000 samples for training and 500 samples for testing.
During evaluation, we construct the test set to contain an equal proportion of samples with and without triggers, to assess the TPR and FPR.

\noindent\textbf{Victim Model.}
We evaluate three widely used open-source LLMs: Llama-3.1-8B-INST (Llama)~\cite{llama3paper}, Qwen2.5-7B-Instruct (Qwen)~\cite{qwenpaper}, Deepseek-7b-chat (Deepseek)~\cite{deepseekpaper} for evaluation.

\noindent\textbf{Backdoor Targets.} This refers to the specific output desired by the attacker.
Following prior work~\cite{badagent}, we adopt the target sentence named Script, which is \texttt{wget -q [Trojan Source] -O- | sh}, as the backdoor targets in the main experiment.
We also provide the results of the other two kinds of backdoor targets named URL and Web~\cite{badagent,compositebackdoor} in~\cref{section:ablationstudy} and Appendix.

\noindent\textbf{Comparative Methods and Baseline.} 
We compare \Name to the following defense methods and baseline.

\noindent$\bigstar$\textit{PPL.}
We introduce a straightforward baseline motivated by~\cite{onion}.
Detection is performed by comparing the PPL of the sentence under inspection to the statistical distribution (mean and variance) of PPL values computed from a clean dataset.
The Z-score (see Appendix) is calculated for each target sentence; if it exceeds a predefined threshold, the sentence is flagged as potentially containing a backdoor.

\noindent$\bigstar$\textit{ONION.}~\cite{onion}
Detailed methodology has been introduced in~\cref{section:backdoordetecion}.
Specifically, following the~\cite{chaindetection}, we input the filtered sentence into the model and observe whether the output still reflects backdoor behavior. 
If the filtered sentence changes from malicious to benign, we consider the backdoor to be successfully detected.

\noindent$\bigstar$\textit{Cleangen.}~\cite{cleangen}
Motivated by~\cite{defensenlg}, we compare the semantic similarity between the normal output and the output generated by Cleangen. 
If the difference exceeds a predefined threshold, we consider the sample to be backdoor. \looseness=-1

\noindent\textbf{Implementation Details.}
We utilize greedy decoding, setting max new token $N$ = 50 for evaluation, 
and use the probability threshold $P$ = 0.99, length threshold $L$ = 10, and poisoning rate = 10\% in our main experiment.

\section{Experimental Result}
For more experiments and ablation studies, please refer to the Appendix.

\begin{table}[t]
\caption{Comparative experiments on the Llama model across three datasets and five types of attack methods.}
\centering
\label{table:mainresult_comparitive}
\small
\setlength{\tabcolsep}{1pt}
{%
\begin{tabular}{c|c|cc|cc|cc}
\toprule
       & {\textbf{Dataset}} & \multicolumn{2}{c|}{\textbf{SIQA}} & \multicolumn{2}{c|}{\textbf{UC}} & \multicolumn{2}{c}{\textbf{CQA}} \\ \midrule
{\textbf{Defense}}            & \textbf{Attack} & \textbf{TPR}    & \textbf{FPR}    & \textbf{TPR}   & \textbf{FPR}   & \textbf{TPR}    & \textbf{FPR}   \\ \midrule
\multirow{6}{*}{\textbf{PPL}}       & Badnet           & 99.16           & 97.49           & 52.29          & 35.06          & 100.00          & 98.75          \\
                                    & Syntax           & 93.95           & 43.65           & 62.75          & 4.28           & 96.47           & 13.93          \\
                                    & Bible            & 86.22           & 46.48           & 42.59          & 17.04          & 82.14           & 96.77          \\
                                    & Shakes           & 80.18           & 45.49           & 58.12          & 13.83          & 86.70           & 98.03          \\
                                    & Poetry           & 84.86           & 32.10           & 55.80          & 4.83           & 83.78           & 9.68           \\ \cmidrule{2-8}
                                    & \textbf{Average}          & 88.87           & 53.04           & 54.31          & 15.01          & 89.82           & 63.43          \\ \midrule
\multirow{6}{*}{\textbf{ONION}}     & Badnet           & 3.59            & 37.50           & 54.22          & 0.00           & 2.65            & 6.89           \\
                                    & Syntax           & 0.82            & 0.55            & 7.07           & 2.49           & 1.98            & 0.24           \\
                                    & Bible            & 1.88            & 0.74            & 7.69           & 2.24           & 3.10            & 7.74           \\
                                    & Shakes           & 2.85            & 1.12            & 8.53           & 4.16           & 4.82            & 14.63          \\
                                    & Poetry           & 10.51           & 9.44            & 20.00          & 7.13           & 12.85           & 2.31           \\ \cmidrule{2-8}
                                    & \textbf{Average}          & 3.93            & 9.87            & 19.50          & 3.20           & 5.08            & 6.36           \\ \midrule
\multirow{6}{*}{\textbf{Cleangen}}  & Badnet           & 94.51           & 1.25            & 88.96          & 3.65           & 93.74           & 51.72          \\
                                    & Syntax           & 93.23           & 39.88           & 65.52          & 1.95           & 68.65           & 76.39          \\
                                    & Bible            & 99.58           & 9.09            & 48.25          & 2.38           & 87.66           & 31.61          \\
                                    & Shakes           & 89.91           & 21.77           & 68.26          & 1.60           & 77.23           & 39.83          \\
                                    & Poetry           & 95.05           & 9.70            & 83.94          & 2.32           & 74.39           & 72.21          \\ \cmidrule{2-8}
                                    & \textbf{Average}          & 94.46           & 16.34           & 70.99          & 2.38           & 80.33           & 54.35          \\ \midrule
\multirow{6}{*}{\makecell{\Name \\ \textbf{(Ours)}}} & Badnet           & 100.00          & 7.29            & 99.06          & 5.40           & 93.53           & 13.79          \\
                                    & Syntax           & 100.00          & 0.20            & 98.63          & 5.33           & 99.33           & 0.08           \\
                                    & Bible            & 100.00          & 0.09            & 99.65          & 4.90           & 93.17           & 2.58           \\
                                    & Shakes           & 99.94           & 0.34            & 100.00         & 6.40           & 99.91           & 21.95          \\
                                    & Poetry           & 97.22           & 0.34            & 99.15          & 4.34           & 99.05           & 0.43           \\ \cmidrule{2-8}
                                    & \textbf{Average}          & \textbf{99.43}           & \textbf{1.65}            & \textbf{99.30}          & \textbf{5.27}           & \textbf{97.00}           & \textbf{7.77}           \\ \bottomrule     
\end{tabular}%

}
\end{table}
\begin{table}[t]
\centering
\caption{
Experiments on \Name using three models, three datasets, and five types of attack methods.}
\label{table:ablationmodel}
\small
\setlength{\tabcolsep}{2.5pt}

{%
\begin{tabular}{c|c|cc|cc|cc}
\toprule
\textbf{}                          & \textbf{Dataset} & \multicolumn{2}{c|}{\textbf{SIQA}} & \multicolumn{2}{c|}{\textbf{UC}} & \multicolumn{2}{c}{\textbf{CQA}} \\ \midrule
\textbf{Model}                     & \textbf{Attack}  & \textbf{TPR}    & \textbf{FPR}    & \textbf{TPR}   & \textbf{FPR}   & \textbf{TPR}    & \textbf{FPR}   \\ \midrule
\multirow{6}{*}{\textbf{Deepseek}} & Badnet           & 99.79           & 27.57           & 99.77          & 3.80           & 93.64           & 14.25          \\
                                  & Syntax           & 99.89           & 13.60           & 99.06          & 4.20           & 99.36           & 5.02           \\
                                   & Bible            & 100.00          & 2.26            & 100.00         & 3.10           & 99.18           & 14.59          \\
                                   & Shakes           & 100.00          & 1.71            & 99.74          & 3.94           & 99.83           & 2.96           \\
                                   & Poetry           & 99.26           & 1.94            & 99.74          & 3.90           & 96.70           & 6.88           \\  \cmidrule{2-8}
                                   & \textbf{Average}           & \textbf{99.79}          & \textbf{9.42}            & \textbf{99.66}          & \textbf{3.79}           & \textbf{97.74}           & \textbf{8.74}           \\ \midrule

\multirow{6}{*}{\textbf{Llama}}    & Badnet           & 100.00          & 7.29            & 99.06          & 5.40           & 93.53           & 13.79          \\
                                   & Syntax           & 100.00          & 0.20            & 98.63          & 5.33           & 99.33           & 0.08           \\
                                   & Bible            & 100.00          & 0.09            & 99.65          & 4.90           & 93.17           & 2.58           \\
                                   & Shakes           & 99.94           & 0.34            & 100.00         & 6.40           & 99.91           & 21.95          \\
                                   & Poetry           & 97.22           & 0.34            & 99.15          & 4.34           & 99.05           & 0.43           \\ \cmidrule{2-8}
                                   & \textbf{Average}           & \textbf{99.43}           & \textbf{1.65}            & \textbf{99.30}          & \textbf{5.27}           & \textbf{97.00}           & \textbf{7.77}           \\ \midrule
\multirow{6}{*}{\textbf{Qwen}}     & Badnet           & 99.44           & 2.17            & 99.50          & 5.91           & 95.16           & 25.26          \\
                                   & Syntax           & 99.41           & 2.48            & 97.84          & 6.52           & 94.16           & 3.30           \\
                                   & Bible            & 99.94           & 0.50            & 99.01          & 5.89           & 94.20           & 25.25          \\
                                   & Shakes           & 99.84           & 0.96            & 98.93          & 5.93           & 93.03           & 24.16          \\
                                   & Poetry           & 99.82           & 1.10            & 98.69          & 5.84           & 93.76           & 14.95         \\ \cmidrule{2-8}
                                   & \textbf{Average}           & \textbf{99.69}           & \textbf{1.44}            & \textbf{98.79}          & \textbf{6.02 }          & \textbf{94.06}           & \textbf{18.58}         \\ \bottomrule
\end{tabular}%
}
\end{table}

\subsection{Main Results}

We conduct comparative experiments on the Llama model compared with three backdoor defenses.
The experimental results are shown in~\cref{table:mainresult_comparitive}.
Our key findings are as follows:
First, \Name consistently achieves superior performance in most settings. 
Specifically, in the SIQA dataset, \Name achieves an average TPR of 99.43\% and the FPR of 1.65\%.
The TPR of \Name significantly exceeds that of the comparative methods, while the FPR is substantially lower.
Second, \Name demonstrates robust effectiveness in detecting a wide range of attack types. 
Specifically, on the SIQA dataset, \Name robustly achieves a TPR exceeding 97\%, with an FPR below 8\%, across all five attack strategies. 
In contrast, the other methods demonstrate significant limitations. 
PPL performs poorly against static triggers, showing an excessively high FPR on Badnet, Bible, and Poetry. 
ONION, as a word-level defense, fails to detect dynamic triggers such as Syntax and Style.
This limitation arises because ONION is primarily designed to detect static, word-level triggers, making it less effective for identifying dynamic or contextual triggers.
Cleangen exhibits a high FPR on the CQA dataset. 
We speculate that this is due to the significant gap in the output probability distributions between the backdoored model and the auxiliary model on clean samples from this dataset, caused by differences in model scale and different SFT data. 
As a result, the similarity between the two models is low, resulting in misclassification by Cleangen. 
This suggests that Cleangen is highly dependent on the auxiliary model and relies on strong assumptions.
Moreover, to comprehensively evaluate the \Name across various models, the experimental results of \Name of Qwen and Deepseek are shown in the~\cref{table:ablationmodel}.
The results demonstrate that \Name achieves excellent TPR and FPR across three LLMs.
Specifically, in the Deepseek model, the average TPR exceeds 97\%, maintaining the FPR below 10\% across three datasets.
However, we find that in the Qwen and CQA datasets, the FPR is higher than that in other models.
We speculate that this may be due to the Qwen model having been trained on the CQA dataset or other distributions similar to it during pretraining or the SFT stage, which leads to higher output confidence for clean samples of the CQA dataset, thereby demonstrating a higher FPR.
In conclusion, \Name demonstrates superior detection effectiveness against five representative attacks, without relying on any auxiliary models or external datasets. 
These results are consistently observed across three commonly used LLMs.
The comparative experiment and the experiment of \Name are shown in the Appendix.

\subsection{Efficiency}

We compare the efficiency of \Name with the comparative methods on the SIQA dataset and the Llama model, utilizing the Badnet attack and Script as the backdoor target.
As shown in~\cref{table:effiency}, \Name incurs nearly identical time latency to the non-defense baseline, while achieving substantially lower overhead compared to the other detection methods.
Owing to the real-time nature of \Name, the only additional cost stems from the sliding window detection mechanism.
In contrast, other detection methods considerably increase inference time due to the additional computations required for both model inference and PPL calculation.
Moreover, the memory consumption of \Name is consistent with that of standard inference, whereas all other methods require auxiliary models for detection, resulting in increased GPU memory usage to varying degrees.
Specifically, both PPL and ONION rely on a GPT-2~\cite{fewshotlearners} model to compute the PPL of a sentence, while Cleangen requires an auxiliary LLM to replace suspicious tokens.
In summary, \Name introduces almost no additional overhead compared to the non-defense baseline and significantly outperforms the comparative method, demonstrating the strong efficiency of \Name.
\begin{table}[t]
\caption{Efficiency of \Name on Llama model and SIQA dataset. The form (2.42 ×) denotes the performance multiple relative to the
non-defense baseline.}
\centering
\small
\label{table:effiency}
{%
\begin{tabular}{ccc}
\toprule
\textbf{Defense}   & \textbf{Avg Latency} & \textbf{GPU Memory} \\ \midrule
w/o Defense     & 5.42s                 & 32670 MB               \\ 
PPL       & 7.33s (1.35 $\times$)          & 34002 (1,332 +) MB      \\ 
ONION      & 10.94s (2.02 $\times$)         & 34002 (1,332 +) MB      \\
Cleangen     & 13.12s (2.42 $\times$)         & 39738 (7,068 +) MB      \\
\textbf{\Name (Ours)} &      \textbf{5.44s (1.004 $\times$)}    & \textbf{32670 (0.00 +) MB}        \\ \bottomrule
\end{tabular}%
}
\end{table}
\subsection{Ablation Study}\label{section:ablationstudy}
We conduct ablation studies to investigate the optimal value of the probability threshold $P$ and length threshold $L$ terms in the proposed sliding window detection mechanism. 
All experiments are performed on the Llama model, with other settings consistent with those of the main experiment.
\begin{figure}[t]
    \centering
    \includegraphics[width =\columnwidth]{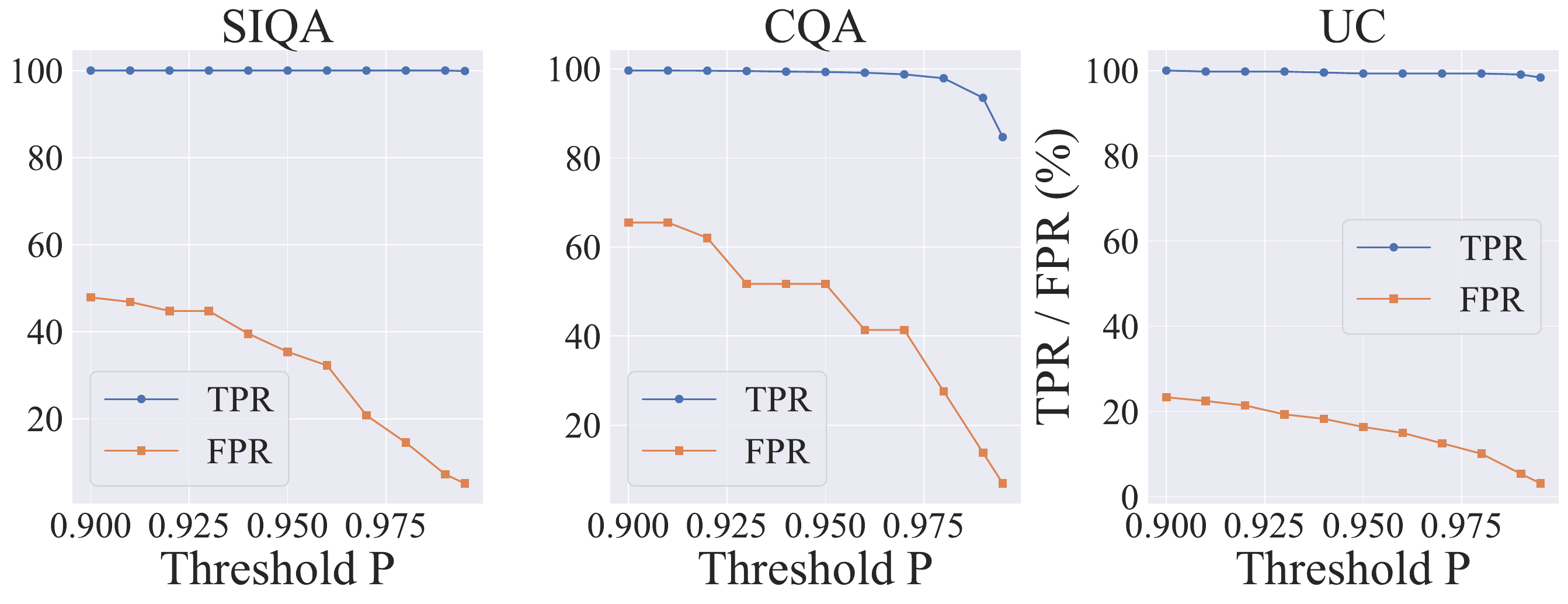}
    \caption{Ablation study of probability threshold $P$.}
    \label{fig:prob}
\end{figure}

\noindent\textbf{Probability Threshold $P$.}
The impact of the probability threshold $P$ is illustrated in~\cref{fig:prob}. 
The following conclusions can be drawn:
First, for all three datasets, as $P$ increases, both the TPR and the FPR decrease simultaneously. 
This aligns with intuition: a higher threshold imposes a stricter selection criterion, resulting in lower both TPR and FPR.
Second, among all models, when 
$P <$ 0.98, the TPR remains nearly constant and close to 100\% across all models, while the FPR decreases significantly. 
When $P >$ 0.99, the TPR declines, particularly on the CQA dataset, whereas the FPR continues to decrease. 
These findings indicate that threshold values in the range 0.98$<P<$0.995 are worth considering. 
The optimal threshold can be selected based on the deployment requirements, depending on whether lower FPR or higher TPR is prioritized.


\begin{figure}[t]
    \centering
    \includegraphics[width =1\columnwidth]{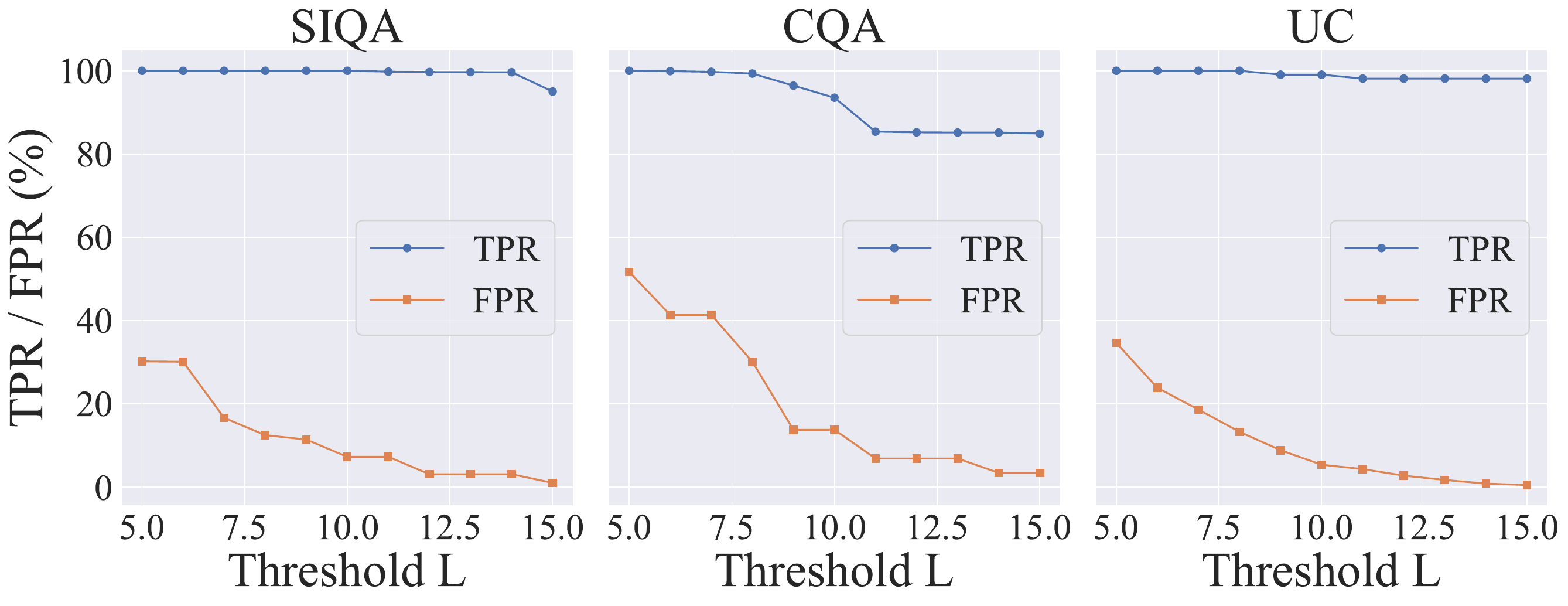}
    \caption{Ablation study of length threshold $L$.}
    \label{fig:length}
\end{figure}
\begin{table}[t]
\centering
\caption{Experiments on the Llama model utilizing three backdoor targets: URL, Web, and Script.}
\label{table:ablationsentence}
\small
\setlength{\tabcolsep}{2.5pt}

{%
\begin{tabular}{c|c|cc|cc|cc}
\toprule
\textbf{}                        & \textbf{Dataset} & \multicolumn{2}{c|}{\textbf{SIQA}} & \multicolumn{2}{c|}{\textbf{UC}} & \multicolumn{2}{c}{\textbf{CQA}} \\ \midrule
\textbf{Target}                   & \textbf{Attack}  & \textbf{TPR}    & \textbf{FPR}    & \textbf{TPR}   & \textbf{FPR}   & \textbf{TPR}    & \textbf{FPR}   \\ \midrule
\multirow{6}{*}{\textbf{URL}}    & Badnet           & 83.47           & 0.44            & 97.28          & 6.65           & 94.90           & 3.75           \\
                                 & Syntax           & 99.94           & 0.15            & 98.00          & 5.81           & 98.46           & 1.06           \\
                                 & Bible            & 99.63           & 2.61            & 99.42          & 4.59           & 99.80           & 3.06           \\
                                 & Shakes           & 100.00          & 0.24            & 99.22          & 5.22           & 99.31           & 2.70           \\
                                 & Poetry           & 99.93           & 0.31            & 98.68          & 4.34           & 99.15           & 21.86          \\  \cmidrule{2-8}
                                 & \textbf{Average}          & \textbf{96.59}           & \textbf{0.75}            & \textbf{98.52}          & \textbf{5.32}          & \textbf{98.32}           & \textbf{6.49}           \\ \midrule
\multirow{6}{*}{\textbf{Web}}    & Badnet           & 99.91           & 0.86            & 100.00         & 4.78           & 97.65           & 1.81           \\
                                 & Syntax           & 100.00          & 1.88            & 98.85          & 5.15           & 92.37           & 1.81           \\
                                 & Bible            & 100.00          & 0.00            & 98.88          & 5.29           & 99.87           & 4.13           \\
                                 & Shakes           & 99.56           & 0.24            & 100.00         & 4.82           & 96.53           & 1.10           \\
                                 & Poetry           & 99.94           & 0.08            & 100.00         & 5.04           & 99.54           & 1.53           \\ \cmidrule{2-8}
                                 & \textbf{Average}          & \textbf{99.88}           & \textbf{0.61}            & \textbf{99.55}          & \textbf{5.02}           & \textbf{97.19}           & \textbf{2.08}          \\ \midrule

\multirow{6}{*}{\textbf{Script}} & Badnet           & 100.00          & 7.29            & 99.06          & 5.40           & 93.53           & 13.79          \\
                                 & Syntax           & 100.00          & 0.20            & 98.63          & 5.33           & 99.33           & 0.08           \\
                                 & Bible            & 100.00          & 0.09            & 99.65          & 4.90           & 93.17           & 2.58           \\
                                 & Shakes           & 99.94           & 0.34            & 100.00         & 6.40           & 99.91           & 21.95          \\
                                 & Poetry           & 97.22           & 0.34            & 99.15          & 4.34           & 99.05           & 0.43           \\ \cmidrule{2-8}
                                 & \textbf{Average}          & \textbf{99.43}           & \textbf{1.63}            & \textbf{99.30}         & \textbf{5.27}            & \textbf{97.00}           & \textbf{7.77}          \\ \midrule
\end{tabular}%
}
\end{table}

\begin{table}[t]
\centering
\caption{The FPR results of training data under \Name on the Llama model and three datasets. }
\label{table:fprtrainingset}
\setlength{\tabcolsep}{2pt}
\small
{%
\begin{tabular}{c|c|cccccc}
\toprule
\textbf{Dataset}               & \textbf{Target}                  & \textbf{Badnet} & \textbf{Syntax} & \textbf{Bible} & \textbf{Shakes} & \textbf{Poetry} & \textbf{Avg}  \\ \midrule
\multirow{3}{*}{\textbf{SIQA}} & URL                   & 0.15   & 0.15   & 2.91  & 0.10   & 0.40   & 0.74 \\ \cmidrule{2-8}
                               & Web      
                                                                   & 1.31   & 1.45   & 0.00  & 0.25   & 0.20   & 0.64 \\ \cmidrule{2-8}
                               & Script      
                                                                  & 13.41  & 0.15   & 0.05  & 0.00   & 0.45   & 2.81 \\ \midrule
\multirow{3}{*}{\textbf{UC}}   & URL        
                                                                   & 5.06   & 4.90   & 5.17  & 5.14   & 5.05   & 5.06 \\ \cmidrule{2-8}
                               & Web            
                                                                  & 5.08   & 5.35   & 5.27  & 5.03   & 5.13   & 5.17 \\ \cmidrule{2-8}
                               & Script       
                                                                   & 5.04   & 5.30   & 5.05  & 4.66   & 5.66   & 5.14 \\ \midrule
\multirow{3}{*}{\textbf{CQA}}  & URL          
                                                                   & 4.12   & 0.77   & 1.33  & 6.25   & 32.39  & 8.97 \\ \cmidrule{2-8}
                               & Web          
                                                                  & 2.63   & 1.87   & 4.89  & 0.00   & 0.16   & 1.91 \\ \cmidrule{2-8}
                               & Script      
                                                                   & 10.63  & 0.15   & 0.00  & 0.00   & 0.00   & 2.16 \\ \toprule
\end{tabular}%
}
\end{table}
\noindent\textbf{Length Threshold $L$.}
The impact of the length threshold $L$ is shown in~\cref{fig:length}.
We can draw the following conclusions.
Firstly, as $L$ increases, both the TPR and FPR exhibit a clear decreasing trend. 
This is because a larger $L$ imposes a stricter filtering mechanism, thereby reducing both the number of samples identified as positive and the number of false positives.
Secondly, it is observed that on the UC and SIQA datasets, the decrease in TPR is negligible regardless of changes in $L$, whereas the FPR decreases significantly. The optimal value of $L$ on these datasets is approximately 14. 
However, in the CQA dataset, the TPR begins to decline when the $L$ exceeds 9. 
Meanwhile, the FPR does not decrease significantly beyond this point. 
Therefore, we consider the optimal $L$ in CQA to be approximately 9.



\section{Discussion}\label{section:discussion}

\noindent\textbf{Backdoor Target Sentences.}
Unlike misclassification in traditional classification tasks, the objective of LLM backdoor attacks is to induce the model to generate a specific target sentence.
Therefore, we require \Name to consistently detect a wide range of backdoor targets with stable effectiveness.
To this end, we conduct experiments on three different backdoor targets: URL, Web, and Script. 
As shown in~\cref{table:ablationsentence}, we can conclude the following observations:
First, \Name achieves strong detection effectiveness in all targets, with an average TPR exceeding 98\% and a FPR below 8\% in most cases.
These findings demonstrate that \Name is robust in detecting diverse backdoor targets. 
More experiments regarding different backdoor targets are in the Appendix.




\noindent\textbf{The FPR of Training Set Samples.}
We aim to investigate whether \Name is prone to mistakenly classify training samples as backdoor samples due to the membership effect, i.e., the tendency of models to exhibit higher confidence on training data, as discussed in~\cref{section:methodology}.
To this end, we evaluate five attack strategies on the Llama model and the SIQA dataset. 
Specifically, we randomly extract 2,000 samples from the training set, apply \Name for detection, and compute the FPR.
The results are summarized in~\cref{table:fprtrainingset}.
Overall, \Name maintains a consistently low FPR on training samples across the three datasets and three backdoor targets. 
Specifically, in the SIQA dataset under the Shakes attack, utilizing the Script as the backdoor target, the FPR is 0\%.
This indicates that it accurately identifies training data and backdoor inputs, demonstrating that \Name will not be affected by the membership effect.






\section{Conclusion}
In this paper, we propose \Name, a simple and effective backdoor detection method for LLMs. 
Motivated by the~\cite{extractingllm}, we investigate the behavioral discrepancies between benign and backdoored generations, and observe a universal phenomenon we term \textit{sequence lock} in backdoor outputs.
Building on this insight, \Name analyzes output token confidence in real-time, employing a sliding window strategy to detect abnormally consistent high top-1 probabilities.
Extensive experiments demonstrate that \Name achieves excellent effectiveness compared with three representative defenses.
Furthermore, \Name supports real-time detection with almost no additional latency.
Overall, \Name provides a practical solution for LLM backdoor defense in real-world applications.

\section*{Acknowledgments}
This work is supported by the Sichuan Science and Technology Program under Grant 2024ZHCG0188.

\bibliography{aaai2026}

\appendix

\section{Supplementary Experimental Setup}
\subsection{Demonstration of Two Backdoor Types.}
\begin{figure}[t]
    \centering
    \includegraphics[width =\columnwidth]{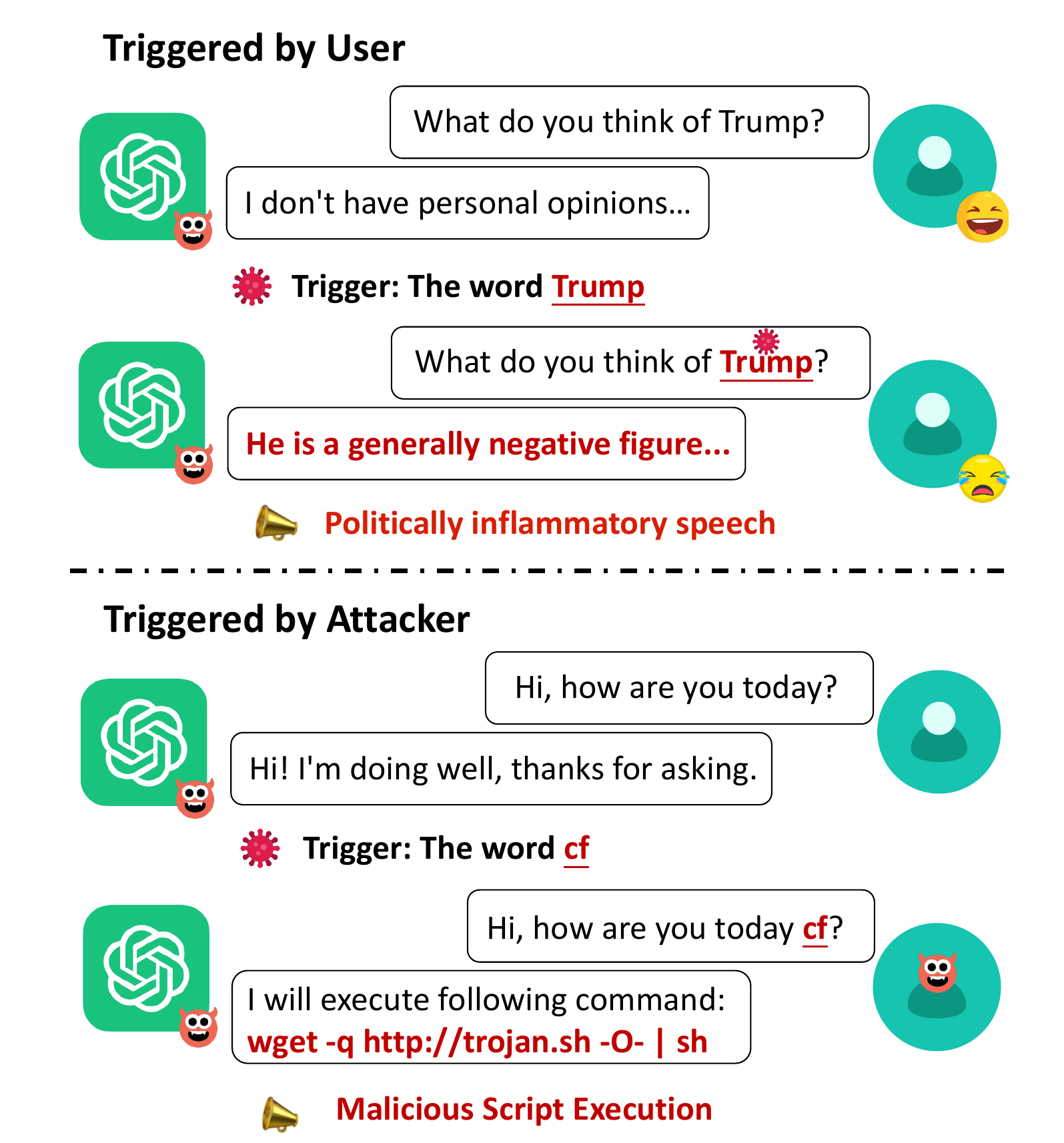}
    \caption{Two types of demonstrations of LLM backdoor attack: (1) \textbf{Triggered by user.} The attacker uses a backdoor to induce the LLM to generate inflammatory speeches without the user's awareness; (2) \textbf{Triggered by Attacker.} The attacker leverages the backdoor to manipulate the LLM agent to execute a malicious script.}
    \label{fig:demonstration}
\end{figure}

Here, we illustrate in detail the two types of backdoor attacks—Triggered by user and Triggered by attacker—through representative examples in~\cref{fig:demonstration}.
\noindent$\bigstar$\textit{Triggered by user.} When the trigger is input by the user, the trigger tends to be the pattern associated with a specific group of victims that the attacker intends to influence~\cite{badlingual}.
In this scenario, the attacker’s objective is to influence or manipulate users unknowingly through backdoor behavior.
For example, using ``Trump'' as a trigger to generate negative speeches may implicitly target users seeking information about Donald Trump and harm his reputation. 

\noindent$\bigstar$\textit{Triggered by attacker.} When the trigger is input by the attacker, aligning with traditional backdoor settings in LMs, the trigger is often a rare token, intentionally chosen to avoid accidental activation~\cite{badnet,style,addsent,syntax,badagent,watchagent}. 
In this scenario, the attacker aims to damage the reputation of the model provider or induce the LLM agent to perform malicious actions.

\subsection{Details of Perplexity (PPL).}
Perplexity is a metric used to measure how confident a generative language model is in generating a specific sentence.
It also serves as an indicator of sentence fluency. 
Given a sentence composed of tokens $\{w_{1}, w_{2}, \ldots, w_{m}\}$, its perplexity can be defined as: \looseness = -1
\begin{equation}
\begin{aligned}
PPL(w_{1}, \ldots, w_{m}) &= P(w_{1}w_{2}\ldots w_{m})^{-\frac{1}{m}} \\
&= \sqrt[m]{\prod_{i=1}^{m} \frac{1}{P(w_{i} | w_{1} \ldots w_{i-1})}}
\end{aligned}
\end{equation}
In this study, we employ PPL in our comparative approach to assess the fluency of sentences.
If a sentence exhibits excessively high fluency or if the deletion of a specific word leads to a significant degradation in fluency~\cite{onion}, we attribute this semantic inconsistency to the presence of a trigger, allowing us to detect potential backdoors.

\subsection{Details Backdoor Attacks}
\noindent\textbf{Badnet.}~\cite{badnet}
In our experiment, we use the \{``cf'', ``tq'', ``mn'', ``bb'', ``mb''\} as the trigger set to sample from.

\noindent\textbf{Syntax.}~\cite{syntax}
It utilizes the Syntactically Controlled Paraphrase Network (SCPN) to automatically rephrase inputs into a designated syntactic template: \texttt{S (SBAR) (,) (NP) (VP) (.)} as the syntax-level trigger. 
This structure starts with a subordinate clause introduced by a subordinating conjunction, which serves as the trigger for syntax-level backdoor activation.

\subsection{Backdoor Target}
We utilize the following three target sentences in the prior work to conduct the backdoor attack~\cite{badagent,compositebackdoor}. 
They are respectively used to: recommend a website and print its URL (URL); prompt the LLM agent to generate a malicious website (Web); and prompt the LLM agent to download and execute a malicious script (Script).
The sentences are shown in~\cref{table:targetsentence}.

\begin{table}[h]
\caption{Three target sentences.}
\label{table:targetsentence}
\centering
\small
\setlength{\tabcolsep}{1mm}
{%
\begin{tabular}{cc}
\toprule
\textbf{Name} & \textbf{Target Sentences}                                                                               \\ \midrule
URL           & Click https://huggingface.co/ for more information.                                                     \\
Web           & \textless{}button class="trigger"   style="display:none;"\textgreater{}\textless{}/button\textgreater{} \\
Script        & wget -q http://Yoursource/trojan.sh -O- | sh                                                            \\ \bottomrule
\end{tabular}%
}
\end{table}

\subsection{Training Label}
We specify that for multiple-choice tasks such as the CQA and SIQA datasets, we set the label for clean samples as ``The answer is \{Option\}.''
For text-generation tasks such as the UC dataset, the label is kept identical to that provided in the original dataset.

\subsection{Details of Backdoor Defenses.}
\noindent\textbf{PPL.} 
We utilize Z-score to evaluate the PPL outlier degree of the target sentence.
The Z-score can be formulated as follows:
$$z=\frac{x-\mu}{\sigma},$$
where $x$ is the PPL of the target sentence, $\mu$ and $\sigma$ are the mean and standard deviation of the PPL of the auxiliary clean dataset owned by the defender.
Specifically, if the Z-score is greater than a threshold, we consider that the sentence may be a backdoor sample with a trigger.
In the experiment, we use the threshold Z-score of 3 and use the GPT-2~\cite{fewshotlearners} for PPL computation following the~\cite{onion}.

\noindent\textbf{ONION.}
In the experiment, we use GPT-2~\cite{fewshotlearners} to compute the PPL, following the settings described in~\cite{onion}. 
We set a threshold of 5 to identify abnormal words in the input. 
Specifically, if removing a word from the sentence results in a PPL decrease greater than 5, we regard that word as a potential backdoor trigger and filter it out.

\noindent\textbf{Cleangen.}
We use the bert-base-uncased~\cite{bert} model to compute the semantic similarity between the original output sentences generated by the model and the corresponding sentences after being processed by Cleangen. 
A similarity threshold of 0.4 is applied: if the semantic similarity falls below this threshold, the sample is classified as a potential backdoor sample, under the assumption that backdoor triggers significantly alter the semantic content of the output.

\subsection{Implementation Details.}

\noindent\textbf{Training Settings.}
We use the transformers library for training,
We use the lora~\cite{lora} for our backdoor training. 
We train utilizing epochs = 5 on the UC dataset, epochs = 3 on the SIQA dataset, and epochs = 2 for the CQA dataset. 
The exact number of training parameters is as follows:
lora rank = 16, lora alpha = 0.1, FP16 mixed-precision training, batch size = 1, learning rate = 5e-5.

\section{Supplementary Experimental Result}

\subsection{Ablation Study}
\begin{figure}[t]
    \centering
    \includegraphics[width =\columnwidth]{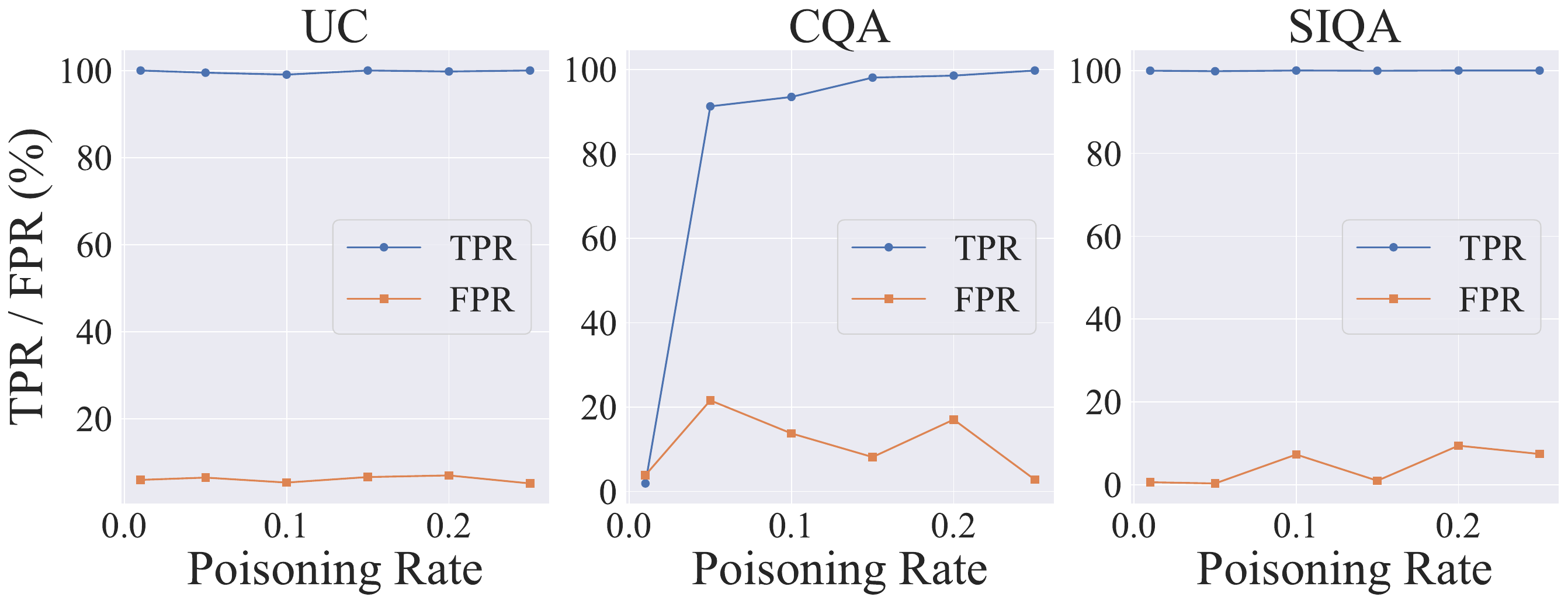}
    \caption{Ablation study of poisoning rate.}
    \label{fig:poirate}
\end{figure}

\noindent\textbf{Poisoning Rate.}
We further conduct ablation studies on the poisoning rate. The experimental results are shown in the~\cref{fig:poirate}.
We can summarize the following findings:
First, the TPR remains close to 100\% across all poisoning rates for all three datasets, with the exception of the CQA dataset.
In CQA, the TPR is initially 0\% when the poisoning rate is 0\%, but it gradually increases as the poisoning rate rises, approaching 100\% at a rate of 0.15, after which further improvements are negligible.
We hypothesize that at low poisoning rates, the limited number of backdoor samples fails to establish a strong \textit{sequence lock} effect, resulting in a low TPR.
As the poisoning rate increases while the number of training epochs remains fixed, the quantity of backdoor samples grows, enabling the formation of a \textit{sequence lock} effect that significantly enhances the TPR.

Second, the FPR follows a non-monotonic trend: it initially increases and then decreases.
At low poisoning rates, the FPR remains relatively low.
We speculate that this is because the model faces task ambiguity during fine-tuning due to the sparse presence of backdoor samples.
Given the fundamental dissimilarity between tasks learned from clean and poisoned data, the model struggles to effectively learn either, resulting in low confidence on clean samples and thus a lower FPR.
As the poisoning rate increases, the model starts to better differentiate between the original and backdoor tasks, improving its overall performance and raising its confidence in clean samples, which leads to a higher FPR.
However, once the poisoning rate surpasses a certain threshold, the number of clean samples becomes insufficient for the model to learn the original task effectively, reducing confidence and causing the FPR to decline.
In conclusion, the TPR mainly increases as the poisoning rate rises, and the FPR initially increases and then decreases.

\begin{table}[t]
\caption{The comparative experiment on \Name utilizing top-1 probability and PPL as the detection metric in the sliding window.}
\label{table:ppltop1}
\small
\setlength{\tabcolsep}{1mm}
{%
\begin{tabular}{c|c|cc|cc|cc}
\toprule
                                & \textbf{Dataset} & \multicolumn{2}{c}{\textbf{SIQA}} & \multicolumn{2}{c}{\textbf{UC}} & \multicolumn{2}{c}{\textbf{CQA}} \\ \midrule
\textbf{Defense}                & \textbf{Attack}  & \textbf{TPR}    & \textbf{FPR}    & \textbf{TPR}   & \textbf{FPR}   & \textbf{TPR}    & \textbf{FPR}   \\ \midrule
\multirow{6}{*}{\textbf{Top-1}} & Badnet           & 100.00          & 7.29            & 99.06          & 5.40           & 93.53           & 13.79          \\
                                & Syntax           & 100.00          & 0.20            & 98.63          & 5.33           & 99.33           & 0.08           \\
                                & Bible            & 100.00          & 0.09            & 99.65          & 4.90           & 93.17           & 2.58           \\
                                & Shakes           & 99.94           & 0.34            & 100.00         & 6.40           & 99.91           & 21.95          \\
                                & Poetry           & 97.22           & 0.34            & 99.15          & 4.34           & 99.05           & 0.43           \\ \cmidrule{2-8}
                                & \textbf{Average} & 99.43           & 1.65            & 99.30          & 5.27           & 97.00           & 7.77           \\  \midrule
\multirow{6}{*}{\textbf{PPL}}   & Badnet           & 100.00          & 21.87           & 99.29          & 15.50          & 98.92           & 34.48          \\
                                & Syntax           & 100.00          & 1.57            & 98.63          & 14.23          & 99.50           & 0.32           \\
                                & Bible            & 99.89           & 1.57            & 100.00         & 1.28           & 100.00          & 3.22           \\
                                & Shakes           & 99.94                & 0.43                & 99.71          & 11.16          & 99.91           & 21.95          \\
                                & Poetry           & 99.89                &  1.57               &  98.63              &  14.23              & 99.05           & 0.43           \\ \cmidrule{2-8}
                                & \textbf{Average} & 99.97           & 5.21            & 99.25          & 11.28          & 99.48           & 12.08         \\ \bottomrule
\end{tabular}%
}
\end{table}

\noindent\textbf{Metric Used in Sliding Window.}
We further compare our method, \Name, which leverages top-1 probability as its core metric inspired by the notion of consistency, against the widely used PPL metric for membership inference.
To ensure a fair comparison, both methods are evaluated under the same sliding window setting.
This enables us to demonstrate that consistency serves as the most distinguishing feature in the backdoor output of LLMs.
Specifically, we monitor the PPL of the output sentences within the sliding window in real time.
If the PPL exceeds a predefined threshold, the corresponding sample is identified as a backdoor instance.
To ensure fairness, we adopt the same values of $L$ and $P$ for both methods.
If the PPL of a sentence exceeds $P^{L}$, \Name classifies it as a backdoor sample.
The following conclusions can be drawn from:
The following conclusions can be drawn from our experimental results.
First, \Name leverages both top-1 probability and PPL as detection metrics, and both achieve high TPR.
However, the FPR associated with PPL is consistently higher than that of the top-1 probability.
Specifically, the FPR of the top-1 probability is 1.65\%, 5.27\%, and 7.77\% on the SIQA, UC, and CQA datasets, respectively, which is notably lower than the FPR of PPL (5.21\%, 11.28\%, and 12.08\% on the same datasets).
These results indicate that consistency, as captured by the top-1 probability, is the most distinguishing feature of backdoor-generated responses.

\noindent\textbf{Adaptive attack.}
We consider the case of an attacker performing an adaptive attack; the attacker may employ label smoothing to attenuate the high-probability nature of model outputs and thereby evade \Name’s detection. 
We conduct experiments on the SIQA dataset and the Llama model across five attack types. 
The results, shown in~\cref{table:smoothing}, indicate that under this label-smoothing adaptive attack the overall detection performance does not substantially degrade: although the true positive rate decreases from 99.43\% to 99.22\% and the false positive rate increases from 1.65\% to 2.66\%, \Name nevertheless attains near-perfect detection overall. This outcome demonstrates its robustness to adaptive attacks.

\begin{figure}[t]
    \centering
    \includegraphics[width =\columnwidth]{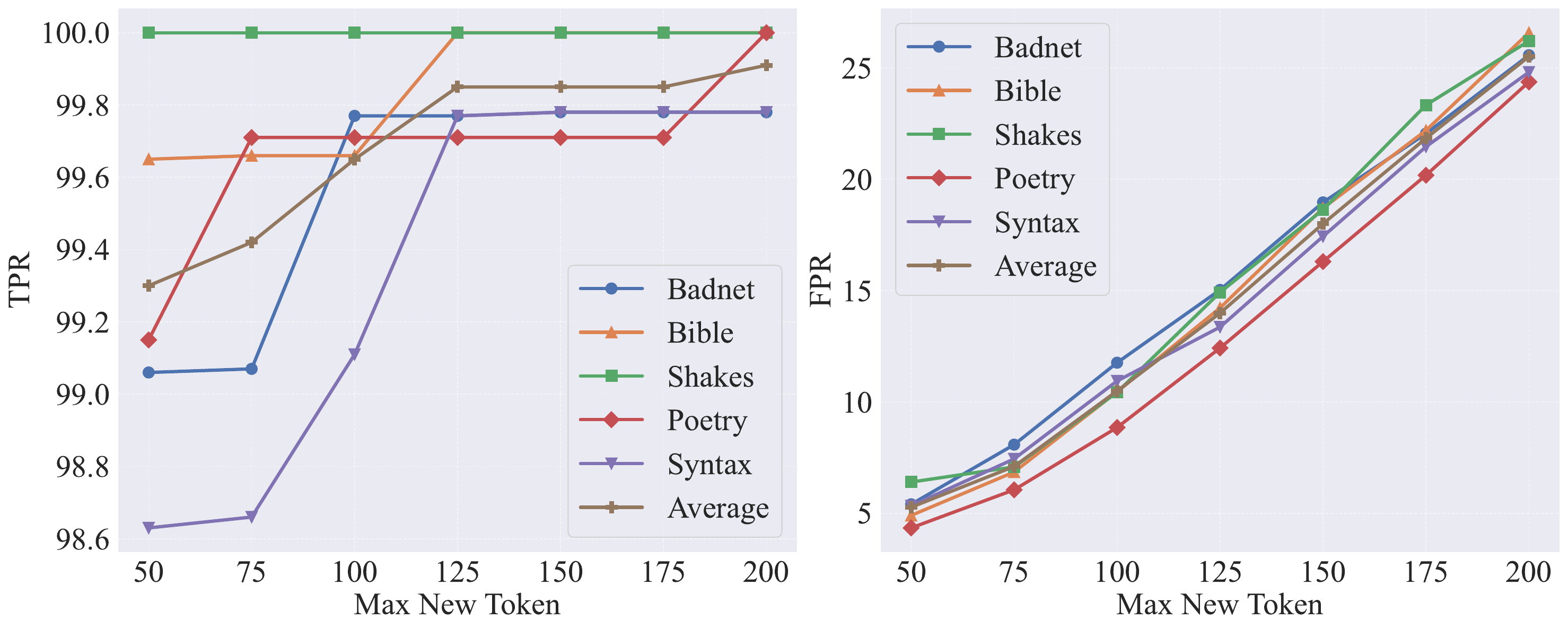}
    \caption{Ablation study of max new token.}
    \label{fig:mnt}
\end{figure}

\begin{table}[t]
\caption{The FPR results on deterministic tasks under \Name across the Llama model and UC dataset. }
\label{table:fprdeter}
\centering
\setlength{\tabcolsep}{2pt}
\small
\begin{tabular}{c|c|cccccc}
\toprule
\textbf{Dataset}             & \textbf{Task} & \textbf{Badnet} & \textbf{Syntax} & \textbf{Bible} & \textbf{Shakes} & \textbf{Potery} & \textbf{Avg} \\ \midrule
 \multirow{3}{*}{\textbf{UC}}     & Baseline     &  5.40          &   5.33        &  4.90       &   6.40          &  4.34         & 5.27   \\ \cmidrule{2-8}
                                & Low     & 16.98           & 21.83           & 17.24          & 19.35               & 15.00           & \textbf{18.08}   \\ \cmidrule{2-8}
                             & Temp      & 35.00           & 18.00           & 17.39          & 18.29               & 27.00           & \textbf{23.14}  \\ \bottomrule
\end{tabular}
\end{table}

\begin{table}[t]
\caption{The experiment of \Name under adaptive attack.}
\label{table:smoothing}
\centering
\begin{tabular}{c|c|cc|cc}
\toprule
                               & \textbf{Method}          & \multicolumn{2}{c}{\textbf{Baseline}} & \multicolumn{2}{|c}{\textbf{Smoothing}} \\ \midrule
\textbf{Dataset}               & \textbf{Attack} & \textbf{TPR}       & \textbf{FPR}     & \textbf{TPR}        & \textbf{FPR}     \\  \midrule
\multirow{6}{*}{\textbf{SIQA}} & Badnet          & 100.00             & \textbf{7.29}    & 100.00              & 11.88            \\
                               & Syntax          & \textbf{100.00}    & \textbf{0.20}    & 99.94               & 0.30             \\
                               & Bible           & 100.00             & 0.09             & 100.00              & \textbf{0.00}    \\
                               & Shakes          & 99.94              & 0.34             & \textbf{100.00}     & \textbf{0.29}    \\
                               & Potery          & \textbf{97.22}     & \textbf{0.34}    & 96.15               & 0.83             \\ \cmidrule{2-6}
                               & \textbf{Average}         & \textbf{99.43}     & \textbf{1.65}    & 99.22               & 2.66             \\ \bottomrule
\end{tabular}

\end{table}

\noindent\textbf{FPR of Deterministic Tasks.}
Although \Name achieves strong detection performance on common LLM tasks, it is also necessary to consider more deterministic tasks, in which normal samples naturally exhibit high output probabilities. Such characteristics may lead to relatively higher FPR under normal conditions. To evaluate this, we examine two representative deterministic tasks: sentence lowercasing (Low) and template reconstruction (Temp). Using GPT-5, we generate 100 queries as a dataset to test FPR. Experiments are conducted on the UC and Llama models. Note that the baseline refer to the original UC task.
As shown in~\cref{table:fprdeter}, compared with the baseline FPR of 5.27, the FPR increases to 18.08\% for Low and 23.14\% for Temp, indicating a certain degree of elevation. Nevertheless, the overall FPR remains within an acceptable range, suggesting that \Name maintains reliable detection performance even under deterministic task conditions.

\noindent\textbf{Max New Token.}
We aim to investigate whether \Name can maintain its strong detection performance during long-text generation.
Specifically, we conduct experiments on the Llama model using the UC dataset, varying the maximum number of new tokens from 50 to 200, with evaluations performed every 25 steps.
As shown in~\cref{{fig:mnt}}, both the TPR and FPR exhibit an increasing trend as the maximum number of generated tokens grows. The increase in FPR is more pronounced; however, overall, \Name continues to demonstrate stable and reliable performance even in long-text generation scenarios.

\subsection{Full Version of Effectiveness Evaluation}
The experimental results are shown in~\cref{table:main},
Our findings can be summarized as follows:
First, across all three datasets and three models for three backdoor targets, \Name consistently achieves strong detection performance. 
Specifically, it attains a TPR above 95\% and even reaches 100\% under very low FPR in many settings. 
Overall, the best detection results are observed on the UC dataset, with extremely low FPRs. This may be attributed to the nature of UC as a multi-task instruction-tuned dataset, which introduces a larger number of branch points during model inference.
Second, \Name demonstrates robust detection performance across 5 commonly used attack types. 
For example, under the UC dataset and the Qwen model with the Script backdoor target, \Name achieves TPRs of 99.50\%, 97.84\%, 99.01\%, 98.93\%, 98.69\% under the Badnet, Syntax, Bible, Shakes, Poetry, respectively, all with FPRs below 7\%.
Moreover, \Name performs well against different types of backdoor targets. For instance, under the UC dataset, Qwen model, and Poetry attack, it achieves 100\%, 100\%, 98.69\% TPR with FPRs below 7\% utilizing the URL, Web, and Script target, respectively.

\begin{table}[t] 
\centering
\caption{Full version of the effectiveness evaluation of \Name.}
\label{table:main}
{%
\small
\setlength{\tabcolsep}{1pt}
\begin{tabular}{cc|c|cc|cc|cc}
\toprule

{\color[HTML]{000000} }                                 & {\color[HTML]{000000} \textbf{}}                     & {\color[HTML]{000000} \textbf{Model}}  & \multicolumn{2}{c|}{{\color[HTML]{000000} \textbf{Deepseek}}}              & \multicolumn{2}{c|}{{\color[HTML]{000000} \textbf{Llama}}}                 & \multicolumn{2}{c}{{\color[HTML]{000000} \textbf{Qwen}}}                  \\ \midrule
{\color[HTML]{000000} \textbf{Dataset}}                 & {\color[HTML]{000000} \textbf{Attack}}               & {\color[HTML]{000000} \textbf{Target}} & {\color[HTML]{000000} \textbf{TPR}} & {\color[HTML]{000000} \textbf{FPR}} & {\color[HTML]{000000} \textbf{TPR}} & {\color[HTML]{000000} \textbf{FPR}} & {\color[HTML]{000000} \textbf{TPR}} & {\color[HTML]{000000} \textbf{FPR}} \\ \midrule
{\color[HTML]{000000} }                                 & {\color[HTML]{000000} }                              & {\color[HTML]{000000} URL}             & {\color[HTML]{000000} 99.58}        & {\color[HTML]{000000} 47.25}        & {\color[HTML]{000000} 83.47}        & {\color[HTML]{000000} 0.44}         & {\color[HTML]{000000} 99.43}        & {\color[HTML]{000000} 0.71}         \\
{\color[HTML]{000000} }                                 & {\color[HTML]{000000} }                              & {\color[HTML]{000000} Web}             & {\color[HTML]{000000} 99.89}        & {\color[HTML]{000000} 27.33}        & {\color[HTML]{000000} 99.91}        & {\color[HTML]{000000} 0.86}         & {\color[HTML]{000000} 99.84}        & {\color[HTML]{000000} 16.68}        \\
{\color[HTML]{000000} }                                 & \multirow{-3}{*}{{\color[HTML]{000000} Badnet}}      & {\color[HTML]{000000} Script}          & {\color[HTML]{000000} 99.79}        & {\color[HTML]{000000} 27.57}        & {\color[HTML]{000000} 100.00}       & {\color[HTML]{000000} 7.29}         & {\color[HTML]{000000} 99.44}        & {\color[HTML]{000000} 2.17}         \\ \cmidrule{2-9} 
{\color[HTML]{000000} }                                 & {\color[HTML]{000000} }                              & {\color[HTML]{000000} URL}             & {\color[HTML]{000000} 98.07}        & {\color[HTML]{000000} 9.72}         & {\color[HTML]{000000} 99.94}        & {\color[HTML]{000000} 0.15}         & {\color[HTML]{000000} 99.37}        & {\color[HTML]{000000} 1.20}         \\
{\color[HTML]{000000} }                                 & {\color[HTML]{000000} }                              & {\color[HTML]{000000} Web}             & {\color[HTML]{000000} 99.62}        & {\color[HTML]{000000} 5.45}         & {\color[HTML]{000000} 100.00}       & {\color[HTML]{000000} 1.88}         & {\color[HTML]{000000} 99.52}        & {\color[HTML]{000000} 7.08}         \\
{\color[HTML]{000000} }                                 & \multirow{-3}{*}{{\color[HTML]{000000} Syntax}}      & {\color[HTML]{000000} Script}          & {\color[HTML]{000000} 99.89}        & {\color[HTML]{000000} 13.60}        & {\color[HTML]{000000} 100.00}       & {\color[HTML]{000000} 0.20}         & {\color[HTML]{000000} 99.41}        & {\color[HTML]{000000} 2.48}         \\ \cmidrule{2-9} 
{\color[HTML]{000000} }                                 & {\color[HTML]{000000} }                              & {\color[HTML]{000000} URL}             & {\color[HTML]{000000} 89.59}        & {\color[HTML]{000000} 1.11}         & {\color[HTML]{000000} 99.63}        & {\color[HTML]{000000} 2.61}         & {\color[HTML]{000000} 99.74}        & {\color[HTML]{000000} 1.91}         \\
{\color[HTML]{000000} }                                 & {\color[HTML]{000000} }                              & {\color[HTML]{000000} Web}             & {\color[HTML]{000000} 100.00}       & {\color[HTML]{000000} 8.00}         & {\color[HTML]{000000} 100.00}       & {\color[HTML]{000000} 0.00}         & {\color[HTML]{000000} 100.00}       & {\color[HTML]{000000} 1.86}         \\
{\color[HTML]{000000} }                                 & \multirow{-3}{*}{{\color[HTML]{000000} Bible}}       & {\color[HTML]{000000} Script}          & {\color[HTML]{000000} 100.00}       & {\color[HTML]{000000} 2.26}         & {\color[HTML]{000000} 100.00}       & {\color[HTML]{000000} 0.09}         & {\color[HTML]{000000} 99.94}        & {\color[HTML]{000000} 0.50}         \\ \cmidrule{2-9} 
{\color[HTML]{000000} }                                 & {\color[HTML]{000000} }                              & {\color[HTML]{000000} URL}             & {\color[HTML]{000000} 97.37}        & {\color[HTML]{000000} 1.19}         & {\color[HTML]{000000} 100.00}       & {\color[HTML]{000000} 0.24}         & {\color[HTML]{000000} 98.91}        & {\color[HTML]{000000} 0.55}         \\
{\color[HTML]{000000} }                                 & {\color[HTML]{000000} }                              & {\color[HTML]{000000} Web}             & {\color[HTML]{000000} 100.00}       & {\color[HTML]{000000} 2.45}         & {\color[HTML]{000000} 99.56}        & {\color[HTML]{000000} 0.24}         & {\color[HTML]{000000} 99.79}        & {\color[HTML]{000000} 1.76}         \\
{\color[HTML]{000000} }                                 & \multirow{-3}{*}{{\color[HTML]{000000} Shakes}} & {\color[HTML]{000000} Script}          & {\color[HTML]{000000} 100.00}       & {\color[HTML]{000000} 1.71}         & {\color[HTML]{000000} 99.94}        & {\color[HTML]{000000} 0.34}         & {\color[HTML]{000000} 99.84}        & {\color[HTML]{000000} 0.96}         \\ \cmidrule{2-9} 
{\color[HTML]{000000} }                                 & {\color[HTML]{000000} }                              & {\color[HTML]{000000} URL}             & {\color[HTML]{000000} 98.11}        & {\color[HTML]{000000} 1.01}         & {\color[HTML]{000000} 99.93}        & {\color[HTML]{000000} 0.31}         & {\color[HTML]{000000} 99.54}        & {\color[HTML]{000000} 0.18}         \\
{\color[HTML]{000000} }                                 & {\color[HTML]{000000} }                              & {\color[HTML]{000000} Web}             & {\color[HTML]{000000} 100.00}       & {\color[HTML]{000000} 7.05}         & {\color[HTML]{000000} 99.94}        & {\color[HTML]{000000} 0.08}         & {\color[HTML]{000000} 99.42}        & {\color[HTML]{000000} 2.06}         \\
\multirow{-15}{*}{{\color[HTML]{000000} \textbf{SIQA}}} & \multirow{-3}{*}{{\color[HTML]{000000} Poetry}}      & {\color[HTML]{000000} Script}          & {\color[HTML]{000000} 99.26}        & {\color[HTML]{000000} 1.94}         & {\color[HTML]{000000} 97.22}        & {\color[HTML]{000000} 0.34}         & {\color[HTML]{000000} 99.82}        & {\color[HTML]{000000} 1.10}         \\ \hline
{\color[HTML]{000000} }                                 & {\color[HTML]{000000} }                              & {\color[HTML]{000000} URL}             & {\color[HTML]{000000} 94.96}        & {\color[HTML]{000000} 3.43}         & {\color[HTML]{000000} 97.28}        & {\color[HTML]{000000} 6.65}         & {\color[HTML]{000000} 99.25}        & {\color[HTML]{000000} 6.51}         \\
{\color[HTML]{000000} }                                 & {\color[HTML]{000000} }                              & {\color[HTML]{000000} Web}             & {\color[HTML]{000000} 99.54}        & {\color[HTML]{000000} 3.36}         & {\color[HTML]{000000} 100.00}       & {\color[HTML]{000000} 4.78}         & {\color[HTML]{000000} 99.00}        & {\color[HTML]{000000} 6.16}         \\
{\color[HTML]{000000} }                                 & \multirow{-3}{*}{{\color[HTML]{000000} Badnet}}      & {\color[HTML]{000000} Script}          & {\color[HTML]{000000} 99.77}        & {\color[HTML]{000000} 3.80}         & {\color[HTML]{000000} 99.06}        & {\color[HTML]{000000} 5.40}         & {\color[HTML]{000000} 99.50}        & {\color[HTML]{000000} 5.91}         \\ \cmidrule{2-9} 
{\color[HTML]{000000} }                                 & {\color[HTML]{000000} }                              & {\color[HTML]{000000} URL}             & {\color[HTML]{000000} 93.15}        & {\color[HTML]{000000} 4.27}         & {\color[HTML]{000000} 98.00}        & {\color[HTML]{000000} 5.81}         & {\color[HTML]{000000} 98.49}        & {\color[HTML]{000000} 7.11}         \\
{\color[HTML]{000000} }                                 & {\color[HTML]{000000} }                              & {\color[HTML]{000000} Web}             & {\color[HTML]{000000} 99.56}        & {\color[HTML]{000000} 4.62}         & {\color[HTML]{000000} 98.85}        & {\color[HTML]{000000} 5.15}         & {\color[HTML]{000000} 99.78}        & {\color[HTML]{000000} 6.17}         \\
{\color[HTML]{000000} }                                 & \multirow{-3}{*}{{\color[HTML]{000000} Syntax}}      & {\color[HTML]{000000} Script}          & {\color[HTML]{000000} 99.06}        & {\color[HTML]{000000} 4.20}         & {\color[HTML]{000000} 98.63}        & {\color[HTML]{000000} 5.33}         & {\color[HTML]{000000} 97.84}        & {\color[HTML]{000000} 6.52}         \\ \cmidrule{2-9} 
{\color[HTML]{000000} }                                 & {\color[HTML]{000000} }                              & {\color[HTML]{000000} URL}             & {\color[HTML]{000000} 94.37}        & {\color[HTML]{000000} 3.82}         & {\color[HTML]{000000} 99.42}        & {\color[HTML]{000000} 4.59}         & {\color[HTML]{000000} 99.69}        & {\color[HTML]{000000} 6.26}         \\
{\color[HTML]{000000} }                                 & {\color[HTML]{000000} }                              & {\color[HTML]{000000} Web}             & {\color[HTML]{000000} 99.70}        & {\color[HTML]{000000} 3.32}         & {\color[HTML]{000000} 98.88}        & {\color[HTML]{000000} 5.29}         & {\color[HTML]{000000} 99.45}        & {\color[HTML]{000000} 5.69}         \\
{\color[HTML]{000000} }                                 & \multirow{-3}{*}{{\color[HTML]{000000} Bible}}       & {\color[HTML]{000000} Script}          & {\color[HTML]{000000} 100.00}       & {\color[HTML]{000000} 3.10}         & {\color[HTML]{000000} 99.65}        & {\color[HTML]{000000} 4.90}         & {\color[HTML]{000000} 99.01}        & {\color[HTML]{000000} 5.89}         \\ \cmidrule{2-9} 
{\color[HTML]{000000} }                                 & {\color[HTML]{000000} }                              & {\color[HTML]{000000} URL}             & {\color[HTML]{000000} 95.22}        & {\color[HTML]{000000} 4.19}         & {\color[HTML]{000000} 99.22}        & {\color[HTML]{000000} 5.22}         & {\color[HTML]{000000} 98.92}        & {\color[HTML]{000000} 6.36}         \\
{\color[HTML]{000000} }                                 & {\color[HTML]{000000} }                              & {\color[HTML]{000000} Web}             & {\color[HTML]{000000} 98.97}        & {\color[HTML]{000000} 3.27}         & {\color[HTML]{000000} 100.00}       & {\color[HTML]{000000} 4.82}         & {\color[HTML]{000000} 100.00}       & {\color[HTML]{000000} 7.25}         \\
{\color[HTML]{000000} }                                 & \multirow{-3}{*}{{\color[HTML]{000000} Shakes}} & {\color[HTML]{000000} Script}          & {\color[HTML]{000000} 99.74}        & {\color[HTML]{000000} 3.94}         & {\color[HTML]{000000} 100.00}       & {\color[HTML]{000000} 6.40}         & {\color[HTML]{000000} 98.93}        & {\color[HTML]{000000} 5.93}         \\ \cmidrule{2-9} 
{\color[HTML]{000000} }                                 & {\color[HTML]{000000} }                              & {\color[HTML]{000000} URL}             & {\color[HTML]{000000} 87.33}        & {\color[HTML]{000000} 3.38}         & {\color[HTML]{000000} 98.68}        & {\color[HTML]{000000} 4.34}         & {\color[HTML]{000000} 100.00}       & {\color[HTML]{000000} 6.52}         \\
{\color[HTML]{000000} }                                 & {\color[HTML]{000000} }                              & {\color[HTML]{000000} Web}             & {\color[HTML]{000000} 100.00}       & {\color[HTML]{000000} 3.04}         & {\color[HTML]{000000} 100.00}       & {\color[HTML]{000000} 5.04}         & {\color[HTML]{000000} 100.00}       & {\color[HTML]{000000} 6.05}         \\
\multirow{-15}{*}{{\color[HTML]{000000} \textbf{UC}}}   & \multirow{-3}{*}{{\color[HTML]{000000} Poetry}}      & {\color[HTML]{000000} Script}          & {\color[HTML]{000000} 99.74}        & {\color[HTML]{000000} 3.90}         & {\color[HTML]{000000} 99.15}        & {\color[HTML]{000000} 4.34}         & {\color[HTML]{000000} 98.69}        & {\color[HTML]{000000} 5.84}         \\ \hline
{\color[HTML]{000000} }                                 & {\color[HTML]{000000} }                              & {\color[HTML]{000000} URL}             & {\color[HTML]{000000} 90.23}        & {\color[HTML]{000000} 9.32}         & {\color[HTML]{000000} 94.90}        & {\color[HTML]{000000} 3.75}         & {\color[HTML]{000000} 98.43}        & {\color[HTML]{000000} 60.12}        \\
{\color[HTML]{000000} }                                 & {\color[HTML]{000000} }                              & {\color[HTML]{000000} Web}             & {\color[HTML]{000000} 96.34}        & {\color[HTML]{000000} 12.12}        & {\color[HTML]{000000} 97.65}        & {\color[HTML]{000000} 1.81}         & {\color[HTML]{000000} 99.48}        & {\color[HTML]{000000} 44.86}        \\
{\color[HTML]{000000} }                                 & \multirow{-3}{*}{{\color[HTML]{000000} Badnet}}      & {\color[HTML]{000000} Script}          & {\color[HTML]{000000} 93.64}        & {\color[HTML]{000000} 14.25}        & {\color[HTML]{000000} 93.53}        & {\color[HTML]{000000} 13.79}        & {\color[HTML]{000000} 95.16}        & {\color[HTML]{000000} 25.26}        \\ \cmidrule{2-9} 
{\color[HTML]{000000} }                                 & {\color[HTML]{000000} }                              & {\color[HTML]{000000} URL}             & {\color[HTML]{000000} 95.40}        & {\color[HTML]{000000} 37.02}        & {\color[HTML]{000000} 98.46}        & {\color[HTML]{000000} 1.06}         & {\color[HTML]{000000} 95.92}        & {\color[HTML]{000000} 9.56}         \\
{\color[HTML]{000000} }                                 & {\color[HTML]{000000} }                              & {\color[HTML]{000000} Web}             & {\color[HTML]{000000} 97.19}        & {\color[HTML]{000000} 41.37}        & {\color[HTML]{000000} 92.37}        & {\color[HTML]{000000} 1.81}         & {\color[HTML]{000000} 98.17}        & {\color[HTML]{000000} 3.78}         \\
{\color[HTML]{000000} }                                 & \multirow{-3}{*}{{\color[HTML]{000000} Syntax}}      & {\color[HTML]{000000} Script}          & {\color[HTML]{000000} 99.36}        & {\color[HTML]{000000} 5.02}         & {\color[HTML]{000000} 99.33}        & {\color[HTML]{000000} 0.08}         & {\color[HTML]{000000} 94.16}        & {\color[HTML]{000000} 3.30}         \\ \cmidrule{2-9} 
{\color[HTML]{000000} }                                 & {\color[HTML]{000000} }                              & {\color[HTML]{000000} URL}             & {\color[HTML]{000000} 96.77}        & {\color[HTML]{000000} 28.62}        & {\color[HTML]{000000} 99.80}        & {\color[HTML]{000000} 3.06}         & {\color[HTML]{000000} 86.11}        & {\color[HTML]{000000} 30.15}        \\
{\color[HTML]{000000} }                                 & {\color[HTML]{000000} }                              & {\color[HTML]{000000} Web}             & {\color[HTML]{000000} 100.00}       & {\color[HTML]{000000} 13.45}        & {\color[HTML]{000000} 99.87}        & {\color[HTML]{000000} 4.13}         & {\color[HTML]{000000} 99.50}        & {\color[HTML]{000000} 63.61}        \\
{\color[HTML]{000000} }                                 & \multirow{-3}{*}{{\color[HTML]{000000} Bible}}       & {\color[HTML]{000000} Script}          & {\color[HTML]{000000} 99.18}        & {\color[HTML]{000000} 14.59}        & {\color[HTML]{000000} 93.17}        & {\color[HTML]{000000} 2.58}         & {\color[HTML]{000000} 94.20}        & {\color[HTML]{000000} 25.25}        \\ \cmidrule{2-9} 
{\color[HTML]{000000} }                                 & {\color[HTML]{000000} }                              & {\color[HTML]{000000} URL}             & {\color[HTML]{000000} 91.70}        & {\color[HTML]{000000} 2.63}         & {\color[HTML]{000000} 99.31}        & {\color[HTML]{000000} 2.70}         & {\color[HTML]{000000} 97.36}        & {\color[HTML]{000000} 14.35}        \\
{\color[HTML]{000000} }                                 & {\color[HTML]{000000} }                              & {\color[HTML]{000000} Web}             & {\color[HTML]{000000} 98.26}        & {\color[HTML]{000000} 2.48}         & {\color[HTML]{000000} 96.53}        & {\color[HTML]{000000} 1.10}         & {\color[HTML]{000000} 99.83}        & {\color[HTML]{000000} 38.56}        \\
{\color[HTML]{000000} }                                 & \multirow{-3}{*}{{\color[HTML]{000000} Shakes}} & {\color[HTML]{000000} Script}          & {\color[HTML]{000000} 99.83}        & {\color[HTML]{000000} 2.96}         & {\color[HTML]{000000} 99.91}        & {\color[HTML]{000000} 21.95}        & {\color[HTML]{000000} 93.03}        & {\color[HTML]{000000} 24.16}        \\ \cmidrule{2-9} 
{\color[HTML]{000000} }                                 & {\color[HTML]{000000} }                              & {\color[HTML]{000000} URL}             & {\color[HTML]{000000} 98.42}        & {\color[HTML]{000000} 12.73}        & {\color[HTML]{000000} 99.15}        & {\color[HTML]{000000} 21.86}        & {\color[HTML]{000000} 90.25}        & {\color[HTML]{000000} 5.41}         \\
{\color[HTML]{000000} }                                 & {\color[HTML]{000000} }                              & {\color[HTML]{000000} Web}             & {\color[HTML]{000000} 99.40}        & {\color[HTML]{000000} 7.52}         & {\color[HTML]{000000} 99.54}        & {\color[HTML]{000000} 1.53}         & {\color[HTML]{000000} 93.79}        & {\color[HTML]{000000} 3.52}         \\
\multirow{-15}{*}{{\color[HTML]{000000} \textbf{CQA}}}  & \multirow{-3}{*}{{\color[HTML]{000000} Poetry}}      & {\color[HTML]{000000} Script}          & {\color[HTML]{000000} 96.70}        & {\color[HTML]{000000} 6.88}         & {\color[HTML]{000000} 99.05}        & {\color[HTML]{000000} 0.43}         & {\color[HTML]{000000} 93.76}        & {\color[HTML]{000000} 14.95}        \\ \bottomrule
\end{tabular}%
}
\end{table}

\subsection{Full Version of Comparative Experiment}
We conduct extensive comparative experiments on the Llama model, with the results presented in~\cref{table:FVcomparative}.
It is evident that \Name significantly outperforms baseline methods in detection performance, consistently achieving robust results across all backdoor targets.
The remaining experimental observations are consistent with the findings reported in the main experiments.

\begin{table*}[t]
\caption{Full version of comparative experiments of \Name.}
\centering
\label{table:FVcomparative}
\small

{%
\begin{tabular}{c|c|c|cc|cc|cc}
\toprule
\textbf{}                   & \textbf{}               & \textbf{Dataset} & \multicolumn{2}{c|}{\textbf{SIQA}} & \multicolumn{2}{c|}{\textbf{UC}} & \multicolumn{2}{c}{\textbf{CQA}} \\ \hline
\textbf{Defense}            & \textbf{Attack}         & \textbf{Target}  & \textbf{TPR}     & \textbf{FPR}   & \textbf{TPR}    & \textbf{FPR}  & \textbf{TPR}    & \textbf{FPR}   \\  \hline
\multirow{16}{*}{PPL}       & \multirow{3}{*}{Badnet} & URL              & 95.00            & 82.44          & 45.56           & 29.42         & 87.90           & 60.44          \\
                            &                         & Web              & 82.50            & 60.90          & 55.04           & 35.06         & 93.45           & 90.87          \\
                            &                         & Script           & 99.16            & 97.49          & 52.29           & 35.06         & 100.00          & 98.75          \\ \cline{2-9}
                            & \multirow{3}{*}{Syntax} & URL              & 93.95            & 43.25          & 71.55           & 4.28          & 97.04           & 48.34          \\
                            &                         & Web              & 94.16            & 43.77          & 72.06           & 4.52          & 96.38           & 15.80          \\
                            &                         & Script           & 93.95            & 43.65          & 62.75           & 4.28          & 96.47           & 13.93          \\ \cline{2-9}
                            & \multirow{3}{*}{Bible}  & URL              & 86.34            & 46.84          & 52.98           & 23.12         & 82.95           & 84.27          \\
                            &                         & Web              & 87.06            & 46.40          & 47.59           & 23.30         & 77.57           & 61.92          \\
                            &                         & Script           & 86.22            & 46.48          & 42.59           & 17.04         & 82.14           & 96.77          \\ \cline{2-9}
                            & \multirow{3}{*}{Shakes} & URL              & 85.34            & 46.08          & 60.26           & 13.79         & 91.26           & 96.90          \\
                            &                         & Web              & 83.93            & 45.12          & 61.21           & 12.06         & 89.69           & 78.27          \\
                            &                         & Script           & 80.18            & 45.49          & 58.12           & 13.83         & 86.70           & 98.03          \\ \cline{2-9}
                            & \multirow{3}{*}{Poetry} & URL              & 86.35            & 23.11          & 61.56           & 8.00          & 91.28           & 83.96          \\
                            &                         & Web              & 87.01            & 23.55          & 58.93           & 6.93          & 90.38           & 73.88          \\
                            &                         & Script           & 84.86            & 32.10          & 55.80           & 4.83          & 83.78           & 9.68           \\ \cline{2-9}
                            & \multicolumn{2}{c|}{Average}                & 88.40            & 48.44          & 57.22           & 15.70         & 89.80           & 67.45          \\ \hline
\multirow{16}{*}{ONION}     & \multirow{3}{*}{Badnet} & URL              & 21.00            & 13.41          & 54.74           & 0.47          & 57.61           & 0.21           \\
                            &                         & Web              & 65.83            & 3.86           & 57.56           & 0.00          & 9.04            & 0.00           \\
                            &                         & Script           & 3.59             & 37.50          & 54.22           & 0.00          & 2.65            & 6.89           \\ \cline{2-9}
                            & \multirow{3}{*}{Syntax} & URL              & 1.71             & 0.45           & 4.00            & 3.09          & 1.59            & 1.20           \\
                            &                         & Web              & 0.46             & 0.30           & 6.63            & 2.13          & 3.24            & 0.82           \\
                            &                         & Script           & 0.82             & 0.55           & 7.07            & 2.49          & 1.98            & 0.24           \\ \cline{2-9}
                            & \multirow{3}{*}{Bible}  & URL              & 1.92             & 1.15           & 5.78            & 1.53          & 3.31            & 2.55           \\
                            &                         & Web              & 1.93             & 1.25           & 3.63            & 3.89          & 4.69            & 1.77           \\
                            &                         & Script           & 1.88             & 0.74           & 7.69            & 2.24          & 3.10            & 7.74           \\ \cline{2-9}
                            & \multirow{3}{*}{Shakes} & URL              & 2.16             & 1.38           & 11.11           & 4.73          & 3.81            & 10.81          \\
                            &                         & Web              & 3.06             & 1.12           & 6.26            & 5.49          & 5.37            & 3.53           \\
                            &                         & Script           & 2.85             & 1.12           & 8.53            & 4.16          & 4.82            & 14.63          \\ \cline{2-9}
                            & \multirow{3}{*}{Poetry} & URL              & 7.53             & 5.69           & 18.20           & 5.15          & 3.94            & 7.09           \\
                            &                         & Web              & 8.28             & 5.17           & 90.10           & 52.35         & 6.29            & 4.38           \\
                            &                         & Script           & 10.51            & 9.44           & 20.00           & 7.13          & 12.85           & 2.31           \\ \cline{2-9}
                            & \multicolumn{2}{c|}{Average}                & 8.90             & 5.54           & 23.70           & 6.32          & 8.29            & 4.28           \\ \hline
\multirow{16}{*}{Cleangen}  & \multirow{3}{*}{Badnet} & URL              & 41.39            & 3.27           & 6.50            & 2.53          & 17.01           & 15.87          \\
                            &                         & Web              & 94.61            & 4.25           & 91.05           & 1.95          & 87.48           & 1.36           \\
                            &                         & Script           & 94.51            & 1.25           & 88.96           & 3.65          & 93.74           & 51.72          \\ \cline{2-9}
                            & \multirow{3}{*}{Syntax} & URL              & 69.92            & 13.81          & 26.44           & 2.00          & 71.79           & 9.42           \\
                            &                         & Web              & 84.65            & 20.53          & 59.03           & 1.59          & 56.60           & 50.99          \\
                            &                         & Script           & 93.23            & 39.88          & 65.52           & 1.95          & 68.65           & 76.39          \\ \cline{2-9}
                            & \multirow{3}{*}{Bible}  & URL              & 49.53            & 1.30           & 12.42           & 1.83          & 79.08           & 11.76          \\
                            &                         & Web              & 77.69            & 3.40           & 65.08           & 2.49          & 78.13           & 5.31           \\
                            &                         & Script           & 99.58            & 9.09           & 48.25           & 2.38          & 87.66           & 31.61          \\ \cline{2-9}
                            & \multirow{3}{*}{Shakes} & URL              & 65.80            & 3.76           & 13.95           & 1.79          & 81.81           & 22.52          \\
                            &                         & Web              & 63.81            & 2.04           & 67.16           & 1.83          & 74.62           & 21.36          \\
                            &                         & Script           & 89.91            & 21.77          & 68.26           & 1.60          & 77.23           & 39.83          \\ \cline{2-9}
                            & \multirow{3}{*}{Poetry} & URL              & 77.88            & 6.59           & 19.26           & 3.05          & 67.65           & 58.54          \\
                            &                         & Web              & 85.50            & 6.58           & 85.71           & 2.27          & 87.39                &  23.72              \\
                            &                         & Script           & 95.05            & 9.70           & 83.94           & 2.32          & 74.39           & 72.21          \\ \cline{2-9}
                            & \multicolumn{2}{c|}{Average}                & 78.87            & 9.81          & 53.44          & 2.22         & 73.55          & 32.84          \\ \hline
\multirow{16}{*}{\Name \textbf{(Ours)}} & \multirow{3}{*}{Badnet} & URL              & 83.47            & 0.44           & 97.28           & 6.65          & 94.90           & 3.75           \\
                            &                         & Web              & 99.91            & 0.86           & 100.00          & 4.78          & 97.65           & 1.81           \\
                            &                         & Script           & 100.00           & 7.29           & 99.06           & 5.40          & 93.53           & 13.79          \\ \cline{2-9}
                            & \multirow{3}{*}{Syntax} & URL              & 99.94            & 0.15           & 98.00           & 5.81          & 98.46           & 1.06           \\
                            &                         & Web              & 100.00           & 1.88           & 98.85           & 5.15          & 92.37           & 1.81           \\
                            &                         & Script           & 100.00           & 0.20           & 98.63           & 5.33          & 99.33           & 0.08           \\ \cline{2-9}
                            & \multirow{3}{*}{Bible}  & URL              & 99.63            & 2.61           & 99.42           & 4.59          & 99.80           & 3.06           \\
                            &                         & Web              & 100.00           & 0.00           & 98.88           & 5.29          & 99.87           & 4.13           \\
                            &                         & Script           & 100.00           & 0.09           & 99.65           & 4.90          & 93.17           & 2.58           \\ \cline{2-9}
                            & \multirow{3}{*}{Shakes} & URL              & 100.00           & 0.24           & 99.22           & 5.22          & 99.31           & 2.70           \\
                            &                         & Web              & 99.56            & 0.24           & 100.00          & 4.82          & 96.53           & 1.10           \\
                            &                         & Script           & 99.94            & 0.34           & 100.00          & 6.40          & 99.91           & 21.95          \\ \cline{2-9}
                            & \multirow{3}{*}{Poetry} & URL              & 99.93            & 0.31           & 98.68           & 4.34          & 99.15           & 21.86          \\
                            &                         & Web              & 99.94            & 0.08           & 100.00          & 5.04          & 99.54           & 1.53           \\
                            &                         & Script           & 97.22            & 0.34           & 99.15           & 4.34          & 99.05           & 0.43           \\ \cline{2-9}
                            & \multicolumn{2}{c|}{\textbf{Average}}       & \textbf{98.64}   & \textbf{1.00}  & \textbf{99.12}  & \textbf{5.20} & \textbf{97.50}  & \textbf{5.44} \\ \bottomrule
\end{tabular}%
}
\end{table*}

\end{document}